\documentclass[a4paper,fleqn,usenatbib]{mnras}

\usepackage{graphicx}	
\usepackage{amsmath}	
\usepackage{amssymb}	
\usepackage[T1]{fontenc}
\usepackage{ae,aecompl}
\usepackage{graphicx}
\usepackage{newtxtext,newtxmath}

\title[On the Spin States of HZ Exoplanets Around M Dwarfs]{On the Spin States of Habitable Zone Exoplanets Around M Dwarfs: The Effect of a Near-Resonant Companion}

\author[Vinson \& Hansen]{
	Alec M. Vinson,$^{1}$\thanks{E-mail:vinson@astro.ucla.edu}
	Brad M. S. Hansen$^{1}$
	\\	
	$^{1}$Mani L. Bhaumik Institute for Theoretical Physics, Department of Physics and Astronomy, University of California, Los Angeles, CA, 90095, USA
}

\date{Accepted XXX. Received YYY; in original form ZZZ}

\pubyear{2017}

\begin{document}
	
	\label{firstpage}
	\pagerange{\pageref{firstpage}--\pageref{lastpage}}
	\maketitle

	\begin{abstract}
		
		One longstanding problem for the potential habitability of planets within M dwarf systems is their likelihood to be tidally locked in a synchronously rotating spin state. This problem thus far has largely been addressed only by considering two objects: the star and the planet itself. However, many systems have been found to harbor multiple planets, with some in or very near to mean-motion resonances. The presence of a planetary companion near a mean-motion resonance can induce oscillatory variations in the mean-motion of the planet, which we demonstrate can have significant effects on the spin-state of an otherwise synchronously rotating planet. In particular, we find that a planetary companion near a mean-motion resonance can excite the spin states of planets in the habitable zone of small, cool stars, pushing otherwise synchronously rotating planets into higher amplitude librations of the spin state, or even complete circulation resulting in effective stellar days with full surface coverage  on the order of years or decades. This increase in illuminated area can have potentially dramatic influences on climate, and thus on habitability. We also find that the resultant spin state can be very sensitive to initial conditions due to the chaotic nature of the spin state at early times within certain regimes. We apply our model to two hypothetical planetary systems inspired by the K00255 and TRAPPIST-1 systems, which both have Earth-sized planets in mean-motion resonances orbiting cool stars.
	\end{abstract}
	

	\section{Introduction}
	
	The discovery of many exoplanetary systems in recent years has accelerated the long-standing  interest in the prospect of discovering and characterising habitable planets around other stars. Because of their ubiquity and their longevity, there has been particular interest in determining the frequency with which habitable planets exist around M dwarf stars. Due to the low luminosity of M dwarfs, the habitable zone in such systems is closer to the central star than in direct analogs of our own Solar System. The transit detection method is thus very well suited to detecting exoplanets within the habitable zone of M dwarfs, with larger light curve transit depths and a higher likelihood of the favorable geometry necessary for a transit detection. The overall advantages that M dwarfs present for harboring and discovering habitable exoplanets are such that it is often called the ``small star advantage''. Data from the \textit{Kepler} mission \citep{DC13,DC15,K13} suggest a probability of 95\% confidence that an Earth sized planet within the habitable zone of an M dwarf can be detected through the transit method within 10 pc. 
	
	Such a detection would  represent an attractive opportunity for detailed atmospheric study and characterisation. However, much of this optimism is predicated on the assumption that a similar level of irradiation to that of Earth correlates with habitable conditions and the existence of liquid water at the surface. Indeed, a long known cautionary note holds that such planets would be tidally locked in a synchronous spin state, raising the possibility of extreme temperature differences and strong winds across the planetary surface, which may lessen the similarity to Earth-like conditions \citep{KWR}. This problem may be mitigated by the behaviour of the planetary atmosphere itself -- such as if it has sufficiently strong atmospheric tides \citep{LWMM} or cloud formation near the substellar point \citep{YBFA}.
	
	It is also possible for a planet to get trapped into a non-synchronous spin-orbit resonance \citep{GP66,GP68}, depending on the degree to which the planet deviates from spherical and the functional dependence of the tidal dissipation. This is believed to be the case for Mercury in our own Solar System. In this paper, we investigate the possibility that the tidal locking can also be weakened by purely gravitational effects. The salient point to note is that most prior analyses of this problem have addressed the case of a single planet orbiting the star. Yet, we know that many of the recently discovered planetary systems are (often highly) multiple \citep{L11,FLR}), even when restricting the sample to only M dwarfs \citep{DC13,BJ16}. Thus, gravitational interactions between neighbouring planets may also influence planetary spin states. One proposed way for a neighbouring planet to influence the spin of another is through spin-orbit resonances of the second kind \citep{GP66,GP68}, where interactions cause the same face of the concerned planet to be oriented towards the neighbouring planet at conjunction, given that the concerned planet is sufficiently deformed. However, this requires that the tidal torque contains a significant dependance on the synodic alignment of the planets, which is doubtful.
	
	Instead, we demonstrate that, if a system contains two planets near a mean-motion resonance, then interactions can be sufficiently strong (and sufficiently rapid) to have a significant effect on the spin state. Although secular interactions can also drive changes in planetary properties, these usually occur on longer timescales and suggest that the planetary spin would simply represent a long-term average over the secular oscillations. As we shall show, resonant interactions can drive evolution on timescales sufficiently fast to lead to interactions with the climate system, opening up the possibility of interesting feedback between the spin evolution and climate. However, a mean motion resonance is a special configuration, and thus the model is only applicable to a fraction of the observed  and postulated planetary systems. Nevertheless, such systems are known to exist. Examples include K2-21 \citep{PSC}, which features two sub-Jovian planets near a 5:3 resonance orbiting around an M0 dwarf; Kepler-32 \citep{SJM13}, an M0 star which features 5 planets with low order commensurabilities between neighbours, and TRAPPIST-1 \citep{TRAP1}, with seven planets orbiting a star $0.08\ M_\odot$, including several that appear to be close to first order commensurabilities. Even if such systems represent only a fraction of the M dwarf planetary population, their special properties may make them preferred candidates for detailed study.
	
	Consequently, in this paper, we model the spin state of an Earth-sized planet near a first order mean-motion resonance with another planetary mass companion. We introduce our model and define our parameter space of interest in \S~\ref{Model}. In \S~\ref{Results} we then present a case study of two hypothetical systems inspired by the K00255  and the TRAPPIST-1 systems. These serve to illustrate the kind of behavior we might expect, and we draw more general conclusions in \S~\ref{Discuss}, including the potential for interaction with the planetary climate.
	
	\section{Model}
	\label{Model}
	
	We are principally interested in systems containing rocky, terrestrial-type worlds, orbiting within the habitable zone of small, cool stars. With the expectation that obliquity should quickly erode for planets within such systems \citep{HLB11}, we will consider a system with a planet of mass $m$ orbiting a star of mass $M_*$ with zero obliquity. We must also allow for potential deviations from a spherical figure, and therefore let $A$, $B$, and $C$ be the principal moments of inertia of the planet, with $C$ being the moment about the spin axis, and choose $ A $ to be the moment about the long axis of the planet (axis in the plane of the orbit) such that $B > A$. We let $\theta$ be the angle formed between the long axis of the planet and a stationary line in the inertial frame. In anticipation that the final equilibrium is one close to  synchronous rotation, we define an angle $\gamma \equiv \theta - M$, with $M$ being the mean anomaly of the planet's orbit. Thus, $\gamma$ can be said to represent deviations from a perfectly synchronous spin state, and roughly corresponds to the longitude of the substellar point on the planet (if the orbit is nearly circular). If we assume that the system is near a 1:1 spin-orbit resonance, such that $\dot{\theta} \approx n$, then we obtain the following equation of motion, as shown by \cite{MD} (hereafter MD99)
	
	\begin{equation}
	\ddot{\gamma} + \textnormal{sign}\left[H(e)\right]\frac{1}{2}\omega_S^2 \sin 2\gamma +  \dot{n} = 0
	\label{eqn:SO}
	\end{equation}
	
	where $H(e)$ is a power series in orbital eccentricity $e$, explained further by MD99, and $\omega_S^2 = 3 n^2 \left(\frac{B-A}{C}\right) |H(e)|$. Equation~\ref{eqn:SO} describes the standard model of a single, spinning, planet in a frame rotating with the orbit. This model, along with a component to describe  tidal damping, is the standard way of modelling the spin evolution of such planets, and can incorporate both synchronous and asynchronous equilibria, depending on the form of the function $H$. The most notable application of this model is to describe the trapping of Mercury in the 3:2 spin-orbit resonance \citep{GP66}. This model is also the basis for our analysis, with one crucial distinction. One traditionally sets $\dot{n} = 0$ at this point and describes the libration of the spin state about various possible equilibria defined by the properties of $H(e)$. This is appropriate for a single planet on a Keplerian orbit, and even for planets in multiple systems when the interactions are predominantly secular. However, the interaction of planets near mean motion resonances can yield small variations in $n$ that prove important for our purposes.
	
	In this paper, we allow mean-motion $n$ to vary by considering the consequences of adding a second planet to the system such that the pair lies near a first order mean-motion resonance. We can describe any fluctuations in $n$ due to a planetary companion via the ``Pendulum Model'', also described in detail by MD99. This model describes variations of a resonant angle $\phi = (j + 1) n' - j n - \bar{\omega}$, where $n$ and $n'$ are the orbital frequencies of the planet and perturber respectively, and $\bar{\omega}$ is the planets longitude of periastron. The constant $j$ is the index which denotes the specific choice of resonance.  With this formulation, then, we describe the evolution of the planet in the context of the circular restricted-three body problem, where the outer satellite is on a perfectly circular orbit in the same plane as the inner satellite and does not react to the presence of the inner body. This model is thus most directly applicable when the perturber is substantially larger than the planet itself, but captures the qualitative behaviour in the case of approximately equal mass pairs as well. Thus, the evolution of $n$ is given by 
	
	\begin{equation}
	\dot{n} = - \frac{1}{j} \omega_M^2 \sin \phi
	\end{equation}
	
	and that the motion of $\phi$ is given by
	
	\begin{equation}
	\ddot{\phi} = -\omega_M^2 \sin \phi
	\end{equation}
	
	where $\omega_M^2 = - 3 j^2 C_r n e$ is, for first-order resonances, the angular frequency of oscillations of the resonant argument. $C_r$ is a constant depending on the nature of the perturber and the resonance, given as $C_r = \left(\frac{m'}{m_c}\right)\left(\frac{a}{a'}\right)nf_d$, with $a$ and $a'$ being the semi-major axis of the orbit of the primary and outer perturber, and $f_d$ being a function of the ratio $a/a'$, tabulated in MD99 as $\frac{a}{a'} f_d = -0.7500$ and $-1.5455$ for the 2:1 and 3:2 resonant cases, respectively. We can now use the pendulum model to calculate $\dot{n}$ in Equation \ref{eqn:SO}, from which we obtain the following equation of motion to describe the spin of the planet of interest
	
	\begin{equation}
	\ddot{\gamma} + \textnormal{sign}\left[H(e)\right]\frac{1}{2}\omega_S^2 \sin 2\gamma + \frac{1}{j} \omega_M^2 \sin \phi = 0
	\label{eqn:SO+MMR}
	\end{equation}
	
	We note that Equation \ref{eqn:SO+MMR} describes the behaviour of  a forced pendulum, a model that arises in a variety of contexts. Indeed, this is one of the favorite subjects of dynamical systems studies (e.g. \citet{BB05}), because it can show a range of interesting chaotic behaviors when the two frequencies $\omega_S$ and $\omega_M$ are of similar order. The only difference here relative to the more commonly analysed harmonic forcing case is that our forcing term $\dot{n}$ is itself a pendulum, with $\phi$ being allowed to circulate or librate. Note also that the behaviour of $\phi$ does not depend on $\gamma$, so there is no feedback into the orbit from the spin evolution. 
	
	\subsection{Parameter Choices \label{params}}
	
	Clearly, in order for this new driving force to exert a significant effect on the planetary spin state, the coefficient of the second pendulum term should have a value of similar order to the first, or larger. Therefore, the behaviour of the system will be characterised by the ratio
	
	\begin{equation}
	\frac{2 \omega_M^2}{ j \omega_S^2} = 2 f_d \left( \frac{ j^{5/2} }{j+1} \right)^{2/3} \left( \frac{m'}{M_*} \right) \left( \frac{e/H(e) }{\left(B-A\right)/C} \right)  \label{wrat}
	\end{equation}
	
	where $j$ is an integer which characterises the particular resonance, along with $f_d$, which is a dimensionless value arising from the expansion of the direct part of the disturbing function, as shown in MD99. Let us also make the simplifying assumption that $ H(e) \sim 1$, as would be the case for the analogue of the synchronous equilibrium to lowest-order in $e$. In principle, this system could be trapped into higher order spin states just like Mercury is, but our goal is to examine the consequences of resonant perturbations for the synchronous state. If we set equation~(\ref{wrat}) equal to unity, we can derive an estimate of the kind of perturber mass that is likely to yield deviations from synchronism around a prototypical M~dwarf (assuming a 2:1 mean motion resonance)
	
	\begin{equation}
	m'  \approx \frac{0.33}{ e}  M_\odot \left(\frac{B-A}{C}\right) \left(\frac{M_*}{0.5\ M_\odot}\right)
	\label{omega_ratio}.
	\end{equation}
	
	It is clear that, the more triaxial a planet is, the bigger is the perturber required to have a significant effect  on the planetary spin, and that a smaller perturber is required if eccentricity is large (because this will amplify the oscillations in $\dot{n}$ resulting from the resonant forcing). Thus, an important question is what values of these parameters are likely to be appropriate. 
	
	We take a value of $(B-A)/C \sim 2 \times 10^{-5}$ as our default value, based on the available empirical evidence determined for the Earth, Mars and Mercury \citep{EKE,Margot,Chen15}, which implies a characteristic perturber mass $m' \sim 2.2/e \  M_{\oplus}$. Estimates of the maximum possible triaxiality \citep{ZL17} suggest that a value of order $\sim 10$ times higher is possible, depending on mass and ocean depth. This would require a correspondingly higher value of $m'$. The optimal perturber mass also depends on eccentricity. The eccentricities for low mass planetary systems are quite difficult to measure, but estimates based on transit timing variations \citep{WL13,VEA15,HL16} suggest that the majority have non-zero but small eccentricities, with estimates ranging from $e \sim 0.01-0.1$. Such values are also consistent with the level of eccentricity expected from in situ assembly models \citep{HM13,H15} or the requirement that many escape capture into resonant chains during an extended period of inward migration \citep{GS}. Thus, we will adopt a characteristic value of $e \sim 0.05$, suggesting an optimal perturber mass $m' \sim 44 M_{\oplus}$ for a $0.5 M_{\odot}$ stellar host. Thus, it appears as though, by simple numerical coincidence, the susceptibility of a terrestrial planet to spin perturbations is such that Neptune-mass companions, which are observed in these systems, can drive interesting spin dynamics.

	As an example of an observed system to which this model might apply, consider the Kepler candidate system K00255, which contains a star of mass $M_* = (0.53 \pm 0.06)\ M_\odot$, a confirmed planet K00255.01 of radius $R=(2.51 \pm 0.3)\ R_\oplus$ and orbital period $T = 27.52$ days, and a candidate planet K00255.02 of radius $R = (0.68 \pm 0.08)\ R_\oplus$ and orbital period $T=13.60$ days. This would imply a system near a 2:1 mean-motion resonance. We note that this system is almost perfect for our needs except that the outer planet is closer to the habitable zone than the inner planet, with the inner planet likely being too hot for life. Even so, the system can serve as a useful illustration of the kinds of behavior we may observe for more Earth-like planets. To estimate planetary mass, \cite{L11} and \cite{FLR} suggest a mass-radius relation of $M_p = M_\oplus(R_p/R_\oplus )^{2.06}$ for $R_p > R_\oplus$, which fits well within the solar system.  Adopting this scaling relation for the outer planet of the K00255 system yields a mass of $m' = 7\ M_\oplus$. Based on the values described above, equation~(\ref{wrat}) implies $ 2 \omega_M^2/j \omega_S^2 \sim 0.11$. Although this is less than unity, it is not negligible, implying the possibility of a substantial interaction between the two terms in equation~\ref{eqn:SO+MMR}.
	
	Another recent system of interest in this regard is the TRAPPIST-1 system \citep{TRAP1}, which contains seven transitting planets orbiting a very low mass ($\sim 0.08 M_{\odot}$) M dwarf. Several of the resulting planets form a chain of near resonant neighbouring pairs, and three lie within, or close to, the calculated habitable zone for a star of this mass. Radii and masses (constrained by transit timing variations) suggest that these planets are also rocky, making them an excellent case study for our model.
	
	The above discussion assumes a constant eccentricity. In principle this is a dynamical quantity, which varies under our formalism adopted from MD99 as
	\begin{equation}
	\dot{e} = j_4 C_r \sin \phi
	\end{equation}
	
	We can see that $e$ and $\phi$ then feed back onto each other, with $\omega_M$ being dependent on $e$. We note, however, that to conserve angular momentum, the variations of eccentricity will depend on the variations of the semi-major axis $a$. If $\delta a_{max}$ is small enough, and the mean eccentricity large enough, then $\delta e_{max}$ will also be small.  If we choose our fiducial $e = 0.05$, and then allow $e$ to evolve with $\phi$ over time, we find that $e$ varies by $\sim 20$\% at most for both our fiducial K00255 and TRAPPIST-1 analogue systems, with semi-major axis varying by $\sim 1$\%. With $\omega_M^2 \propto e$, we find that the strength of the forcing term in Equation \ref{eqn:full_motion} can vary by as much as $\sim 20$\% and the driving frequency by $\sim 4$\% due to these variations in eccentricity. Though this can give us slightly different quantitative results for any individual simulation, we find that behavior remains qualitatively similar if we force $e$ to be constant in our simulations. Thus, for our calculations to follow, we will assume a constant $e=0.05$ for clarity.\\
	\\
	
	\subsubsection{Tidal Damping \label{section:circularization}} 
	
	In order for the driving force in our model to affect the spin-state of the primary, eccentricity must have a non-zero value. Although we noted above that there are potential reasons to believe formation may leave small remnant eccentricities, the long term operation of tides can act to damp and circularise eccentricity. This is especially true in compact multiple planet systems, where the effects of secular interactions can extend the reach of tidal circularisation significantly \citep{GvL,HM15}. The empirical estimates based on TTV measurements suggest that many systems have not yet been completely circularised. Nevertheless, it is important to confirm that there is a reasonable expectation of finite eccentricity in these systems. Furthermore, the effect of tides on spin will add another dimension to equation~(\ref{eqn:SO+MMR}), acting to synchronise the spin and thus enforcing the dynamics of a damped, forced pendulum.
	
	The effects of tides on stellar and planetary orbits is a subject of longstanding study (see \cite{Og14} for a review), and the strength of the effects can be calibrated for giant planets (e.g. \cite{H10}). However, the data for lower mass planets is not good enough for a direct calibration, and we will draw on the traditional calibration relative to the strength of tides on Earth. This is traditionally expressed in terms of the `tidal Q' parameter, but we wish to retain the functional form of our prior formulation based on the work of \cite{H81} and \cite{EKH98}, because it has the attractive property that $\ddot{\gamma} \rightarrow 0$ as $\dot{\gamma} \rightarrow 0$. Thus, using equations~(4) and (14) of \cite{H10} we may describe the evolution of $\gamma$ under the action of tides with
	\begin{equation}
	\ddot{\gamma} = -\frac{15}{2} \dot{\gamma} \frac{M_*}{m} \left( \frac{R}{a} \right)^6 M_* R^2 \sigma 		\label{eqn:gamadot}
	\end{equation}
	
	where we express the strength of dissipation in terms of a bulk dissipation constant $\sigma$, and then calibrate that by calculating the equivalent value of $Q$ under the specific forcing applied to Earth. In this expression, $a$ is the semi-major axis of the orbit and $R$ is the planetary radius. 
	
	Under an assumption that the initial spin rate  $\dot{\theta} \gg n$, then for Earth-sized planets in the habitable zone of an M dwarf (e.g. $M_* = 0.5\ M_\odot$), this yields a spin-down (i.e. synchronization) timescale of 
	\begin{equation}
	\tau_{sync} \sim 2\times 10^6\ \textnormal{years} \left(\frac{a}{0.15\ \textnormal{AU}}\right)^6 \left(\frac{R}{R_\oplus}\right)^{-6} \frac{m}{M_\oplus} \left(\frac{M_*}{0.5\ M_\odot}\right)^{-2}\frac{Q_{\oplus}}{10}.
	\end{equation}
	
	We can calculate the circularization timescale in the same formalism, under the assumption that synchronization has happened quickly and $\dot{\theta} \simeq n$, we get
	
	\begin{equation}
	\tau_{circ} =  10^{11} \textnormal{years} \left(\frac{a}{0.15\ \textnormal{AU}}\right)^{13/2} \left(\frac{R}{R_\oplus}\right)^{-5} \frac{m}{M_\oplus} \left(\frac{M_*}{0.5\ M_\odot}\right)^{-3/2}\frac{Q_{\oplus}}{10}. \label{tcirc}
	\end{equation}
	
	This demonstrates that our formulation is consistent, in the sense that planetary spins should be rapidly damped to the system equilibrium, but that orbital eccentricities can persist for the length of system age except for the very shortest period planets. In fact, if we use this formalism to describe  the K00255 system, we get synchronization timescales of $6\times 10^5$ years for the inner planet and $10^6$ years for the outer, while circularization timescales become $2\times 10^{11}$ years and $5.6\times 10^9$ years respectively. At first glance, this is a curious result, as it implies the more distant planet is circularised but the closer one is not. This is because of the strong sensitivity of the tidal strength to planet radius, and serves to further bolster our adoption of the circular restricted problem as a description of the system.
	
	Equation~(\ref{eqn:gamadot}) provides a torque on the spin that will act to damp this libration of $\gamma$. There is a growing literature on the mechanisms of tidal dissipation in earth-like planets \citep{EL07,ME13,FM13,CBLR}, but this simple formalism is sufficient to illustrate the necessary behaviour. In particular, we note that the torque goes to zero as $\dot{\gamma}$ goes to zero, thereby avoiding unphysical discontinuities in the torque at synchronism. Adding this to Equation \ref{eqn:SO}, we find
	
	\begin{equation}
	\ddot{\gamma} + \frac{1}{2} \omega_S^2 \sin 2\gamma + \frac{\omega_M^2}{j} \sin \phi - \epsilon \dot{\gamma}  = 0.
	\label{eqn:full_motion}
	\end{equation}
	
	where $\epsilon = \frac{15}{2}\frac{M_*}{m}\left(\frac{R}{a}\right)^6 M_*R^2\sigma$. We also define a parameter $\beta = 2\omega_M^2/j\omega_S^2$, so that $\beta$ and $\epsilon$ will together characterise the behaviour of $\gamma$. It is the evolution of the spin under the influence of this equation that will determine the final spin state of the planet and the pattern of irradiation that it experiences.
	
	Our model is constructed to consider the evolution of $\gamma$ when near the synchronously rotating resonant case, when $\dot{\gamma}$ is small compared to the mean-motion $n$. However, equilibria potentially exist for $\gamma = \theta - pM$, where $p$ is any integer multiple of $1/2$. One can then perform a similar analysis, with 
		$H(e)$ now also a function of $p$. One example is the Mercury 3:2 spin-orbit case, where the evolution of $\gamma$ is studied choosing $p=3/2$.
	
	Thus, we note that the synchronously rotating case ($p=1$) is not the only spin-state that a planet can be captured into, and other resonant states would in fact result in stable asynchronous rotations. This potentially provides yet another solution to the problem of tidal-locking within the habitable zone of cool stars. The criteria for capture into one of these resonant states depends on the properties of the tidal dissipation\citep{GP66}. Translated into our formalism, this requires the inequality $| \epsilon \dot{\gamma}_{max} | < \frac{1}{2}\omega_S^2$ be satisfied. If this criterion is satisfied for a particular choice in $p$, then one may hope to determine the probability of capture into the corresponding spin-orbit resonance. Our study focusses on just the $p=1$ case, where the planet is very near the synchronously-rotating case, as the default configuration that will result if capture into higher order resonances is not possible.
	
	\section{Results}
	\label{Results}
	
	Equation~(\ref{eqn:full_motion}) will apply to the spin of any terrestrial planet in orbit about a low mass star, with a companion in an orbit close to a mean motion resonance. However, as we have noted, the effects are likely to be strongest when $\omega_S \sim \omega_M$ (i.e. $\beta \sim 1$). As such, we focus here on the hypothetical system based on the properties of K00255. This system contains both an Earth-like interior planet and an external perturber of mass $\sim 7 M_{\oplus}$ near the 2:1 commensurability, in orbit around a 0.53$M_{\odot}$ star. However, the inner planet is too close to the star and lies interior to the nominal habitable zone for this star.
	
	For our illustrative simulations, we therefore choose to place the spinning $1 M_{\oplus}$ planet at a location farther from the star ($0.15 AU$), such that it receives the same stellar flux as Earth, with $M_* = 0.5\ M_\odot$, $e=0.05$, and $\left(B-A\right)/C = 2\times 10^{-5}$. We then place the companion of mass $m'=7\ M_\oplus$ on a perfectly circular outer orbit such that the system forms a 2:1 mean-motion resonance with the primary planet, and examine the spin dynamics of this system. These parameters yield a value of $\beta = 0.11$.
	
	For all following simulations in this paper, we solve Equation~(\ref{eqn:full_motion}) for $\gamma$ using the Runge-Kutta fourth-order method of integration. Time steps adapt to the highest value of $\omega_M$ and $\omega_S$, taking 70 steps in the period corresponding to the highest frequency. To start with, let us examine the evolution of our K00255 analogue system in the absence of dissipation (i.e. $\epsilon=0$).
	
	The dynamics of a pendulum also requires the specifications of initial conditions. In this case, we not only require $\gamma$ and $\dot{\gamma}$, but also $\phi$ and $\dot{\phi}$. Starting with the latter pair, we choose $\phi=0$ initially for all runs, but different $\dot{\phi}$. The value of $\dot{\phi}(0)$ will then characterise the dynamics of the resonant pair -- small enough values will yield libration of $\phi$ and larger values will result in circulation. It is also important to note that, when the driving force is of pendulum form, $\omega_M$ is a characteristic value, but the actual forcing frequency will be a function of $\dot{\phi}(0)$. We must also choose initial conditions in $\gamma$ and $\phi$. We set $\gamma(0) = 0$,  and $\phi(0) = 0$. The choices of $\dot{\gamma}(0)$ and $\dot{\phi}(0)$ will then determine the initial spin state and  the nature of the driving force. 
	
	\begin{figure*}
		\includegraphics[width=\textwidth]{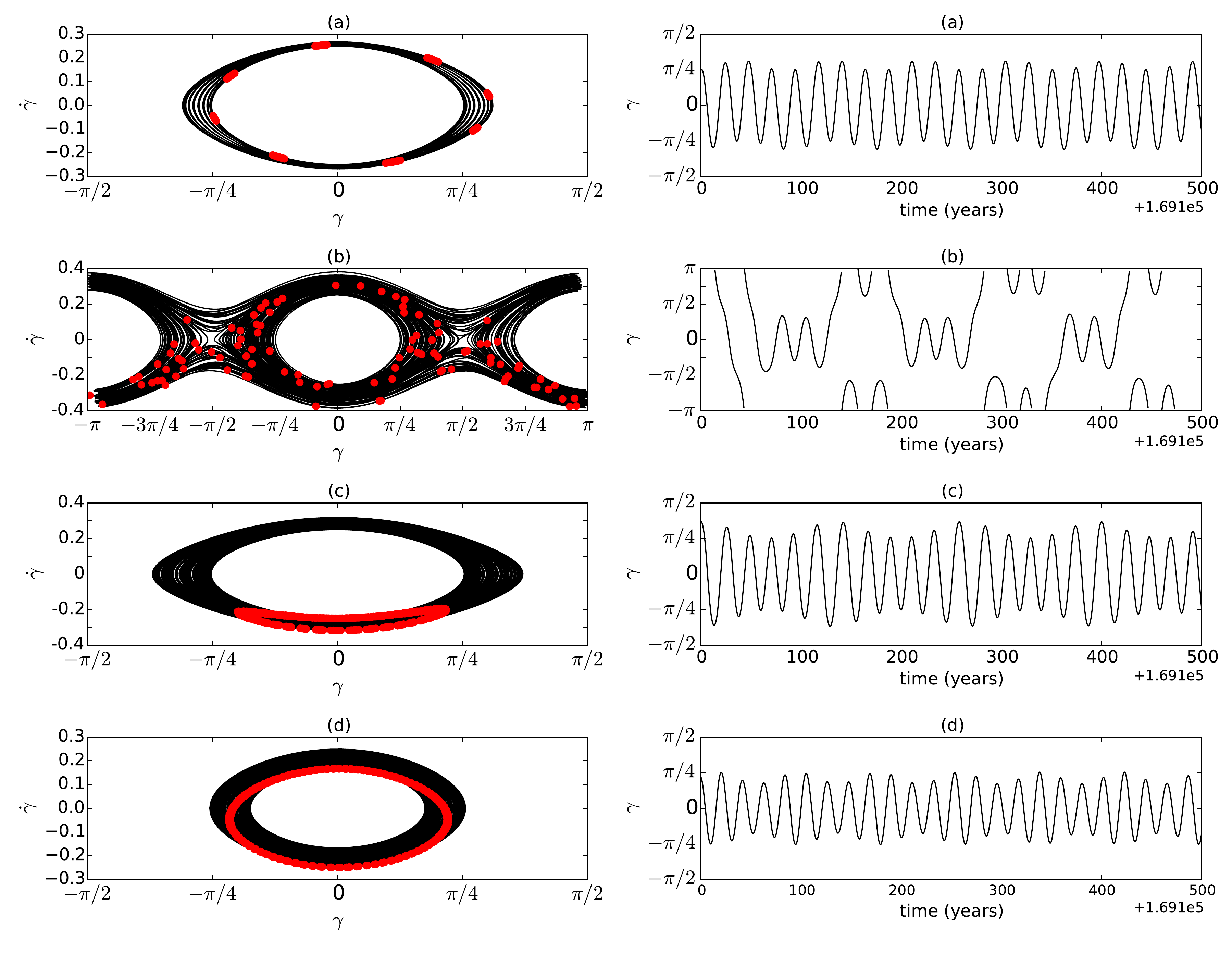}
		\caption[case_comparison.pdf]{The panels show the evolution of $\gamma$ in our K00255 analogue system subject to driving without dissipation ($\epsilon = 0$) according to equation~(\ref{eqn:full_motion}) with $\dot{\phi}(0)=0.09$ yr$^{-1}$ (panels a), $\dot{\phi}(0)=0.21$ yr$^{-1}$ (panels b), $\dot{\phi}(0)=0.29$ yr$^{-1}$ (panels c) and $\dot{\phi}(0)=0.39$ yr$^{-1}$ (panels d). For each simulation, $\dot{\gamma}(0) = 0.25$ yr$^{-1}$, $\gamma (0) = 0$, $\phi(0) = 0$, and $\beta = 0.11$. In each panel on the left, the black curves represent the full late-time evolution, while the red points represent a surface of section taken when $\phi =0$ and $\dot{\phi}>0$. We also show the evolution of $\gamma$ in time on the right.
			\label{Snap1}}
	\end{figure*}
	
	The separatrix of the $\gamma$ -- $\dot{\gamma}$ system occurs at $\dot{\gamma} = 0.33$ yr$^{-1}$ when $\gamma = 0$ for its natural oscillation (no driving force). We therefore consider three reprsentative states of the $\gamma$--$\dot{\gamma}$ spin system consisting of small amplitude libration ($\dot{\gamma}(0) = 0.02$ yr$^{-1}$), large amplitude libration ($\dot{\gamma}(0)=0.25$ yr$^{-1}$), and circulation ($\dot{\gamma}(0)=0.38$ yr$^{-1}$), and examine how each responds to different levels of forcing as determined by the amplitude of $\dot{\phi}(0)$. For all these cases, we let the simulations run for $\sim 10^5$ years to allow sufficient time to approach any final spin-state.
	
\begin{figure}
		\includegraphics[width=\columnwidth]{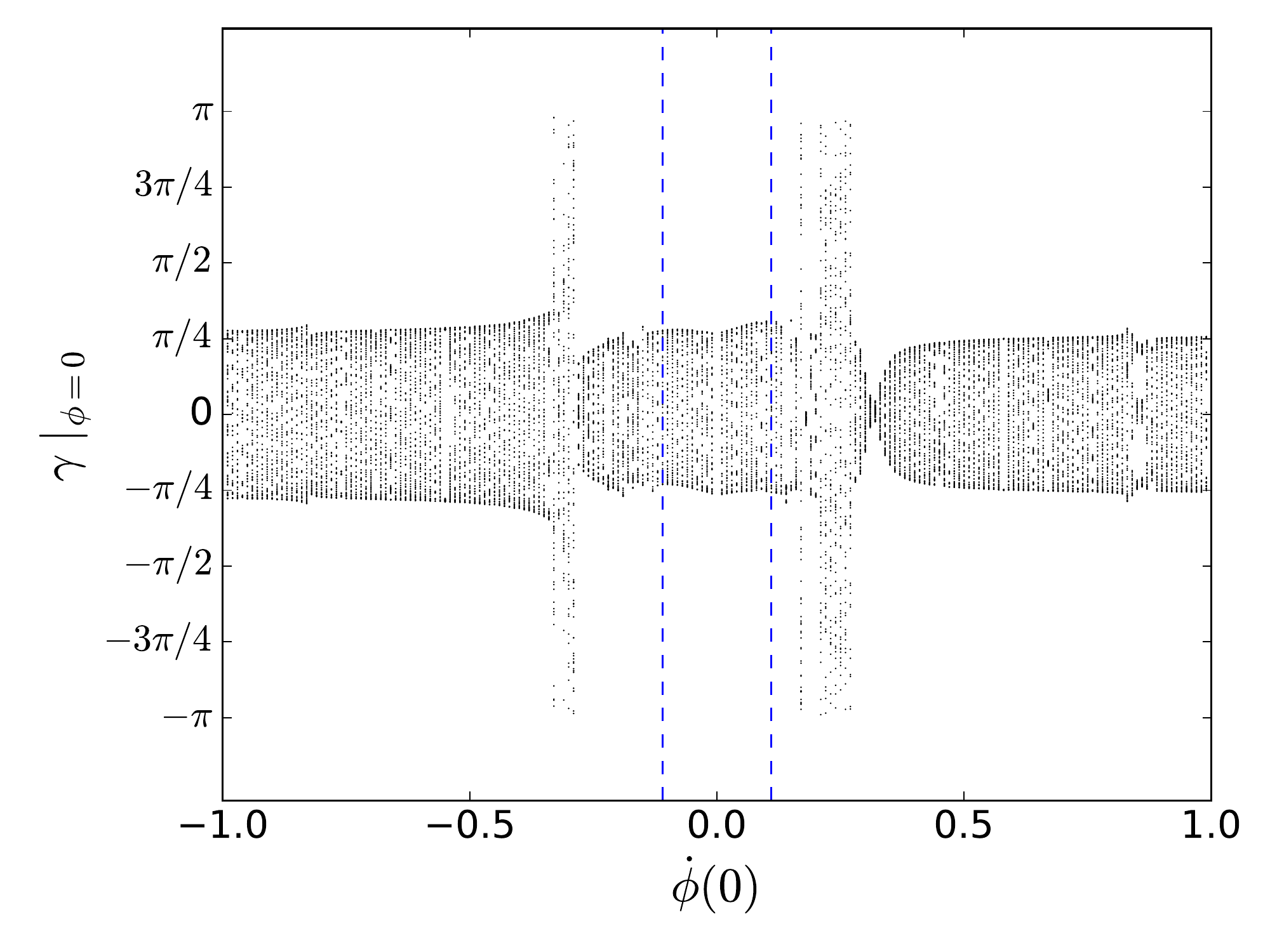}
		\caption[poincare_proj_k255_gp25.pdf]{The points depict the values of $\gamma$ in the surface of section defined by $\phi=0$, and $\dot{\phi}$ with the sign of its initial value, for each value of the initial forcing $\dot{\phi}(0)$ when $\dot{\gamma}(0) = 0.25$ yr$^{-1}$ for our K00255 analogue system, yielding $\beta = 0.11$, in the absence of damping ($\epsilon = 0$). The vertical dashed lines indicate the seperatrix between libration and circulation (to the far left and right) of the external forcing resonant argument. Being in between these dashed lines is thus when the system is {\it in} a mean-motion resonance, with a librating resonant angle $\phi$. We see that the character of the behaviour is the same for both negative and positive values of $\dot{\phi}(0)$, but that it is not entirely symmetric.
			\label{Amp}}
	\end{figure}

	Figure~\ref{Snap1} shows the evolution of the spin for the large amplitude libration case ($\dot{\gamma}(0)=0.25$ yr$^{-1}$) in response to four different levels of forcing. In panel a), $\dot{\phi}(0)=0.09$ yr$^{-1}$, which has little effect. The behaviour seen here is qualitatively the same as if there was no forcing. To better characterise the behaviour, we highlight in red the values of $\gamma$ and $\dot{\gamma}$ that occur when $\phi=0$ and $\dot{\phi}>0$, known as a Poincare section (e.g. \cite{BB05}). In panel b), we show the same evolution, but with $\dot{\phi}(0)=0.21$ yr$^{-1}$. We see here that the spin is now chaotic, switching between circulation and libration about both $\gamma=0$ and $\gamma=\pi$. It is worth noting that $\phi$ is circulating here, since the seperatrix between libration and circulation occurs for $\dot{\phi}(0)=0.11$ yr$^{-1}$ in this case. Thus, the most dramatic effects occur for $\phi$ {\it near} resonance, but not actually in it. In panel (c), we see the behaviour for $\dot{\phi}(0)=0.29$ yr$^{-1}$. Here $\gamma$ is no longer chaotic, but shows a slight broadening of the libration trajectory. We note also that the Poincare section in red produces a closed curve for $\dot{\gamma} < 0$ yr$^{-1}$. This suggests an approximate commensurability between the period of libration for $\gamma$ and period of circulation for $\phi$. As we continue to increase $\dot{\phi}(0)$ this commensurability gets weaker and the surface of section expands, eventually enclosing the origin again, as seen in panel (d), for $\dot{\phi}(0)=0.39$ yr$^{-1}$. We can summarise this behaviour in Figure~\ref{Amp}, which shows the values of $\gamma$ corresponding to the surface of section $\phi=0$, $\dot{\phi}>0$, plotted against initial $\dot{\phi}(0)$.
	
	\begin{figure}
		\includegraphics[width=\columnwidth]{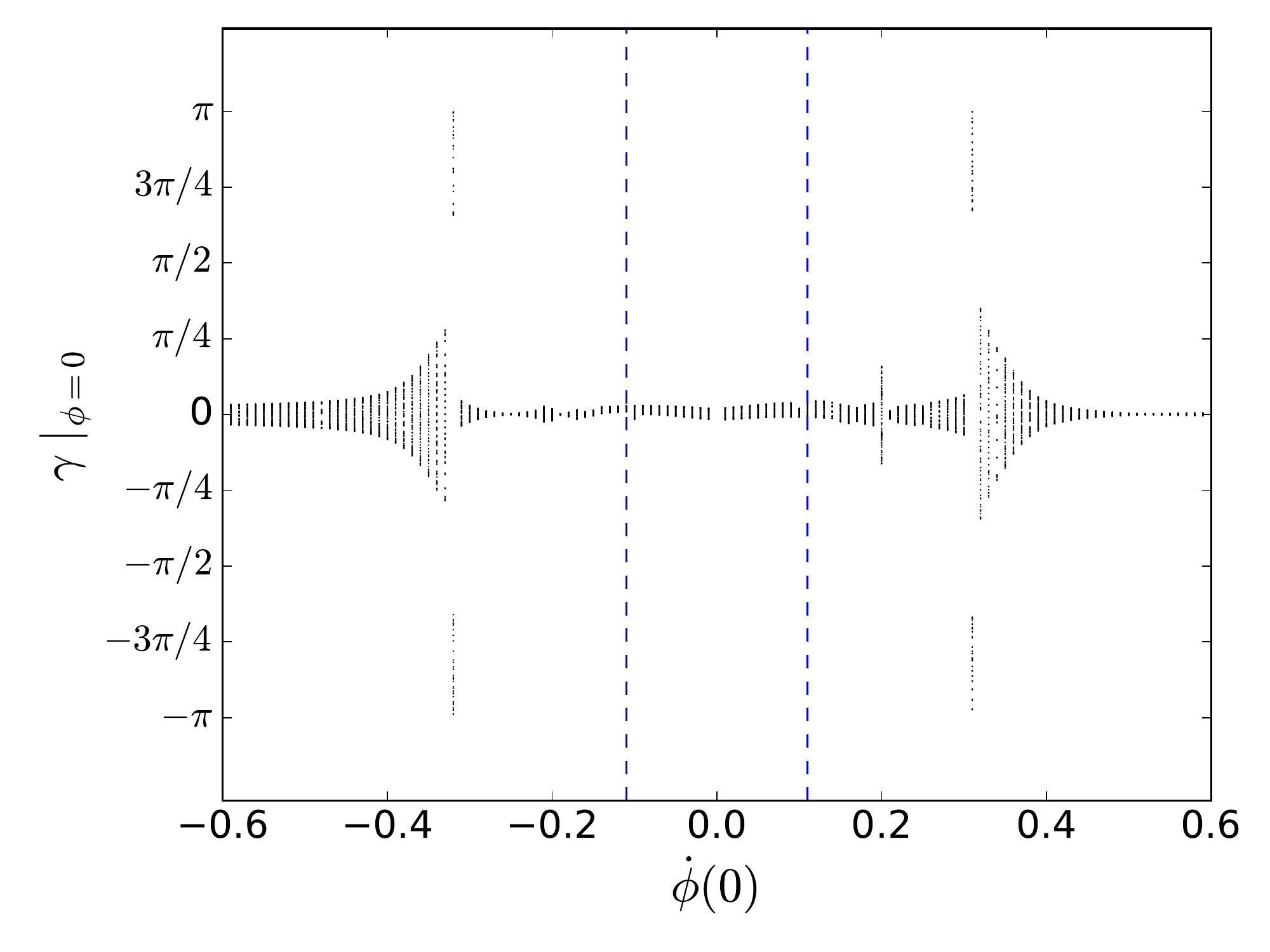}
		\caption{The points depict the values of $\gamma$ in the surface of section defined by $\phi=0$, and $\dot{\phi}$ with the sign of its initial value, for each value of the initial forcing $\dot{\phi}(0)$ when $\dot{\gamma}(0) = 0.02$ yr$^{-1}$ for our K00255 analogue system, yielding $\beta = 0.11$, in the absence of damping ($\epsilon = 0$). The vertical dashed lines indicate the seperatrix between libration and circulation (to the far left and right) of the external forcing resonant argument. Being in between these dashed lines is thus when the system is \textit{in} a mean-motion resonance, with a librating resonant angle $\phi$. We see that the character of the behaviour is the same for both negative and positive values of $\dot{\phi}(0)$, but that it is not perfectly symmetric.
			\label{Amp4}}
	\end{figure}
	
	The large amplitude libration may be especially susceptible to showing chaotic behaviour, since the natural oscillations (in the absence of forcing) are still a non-negligible amplitude relative to the seperatrix value. Thus, in Figure~\ref{Amp4} we consider the effect of a similar range of forcing, but now our initial $\dot{\gamma}(0)=0.02$ yr$^{-1}$, so that the natural libration is of a much smaller amplitude. Once again, the effect of a librating $\phi$ is minimal, with the libration in $\gamma$ not significantly affected, but with a circulating $\phi$, with $\dot{\phi}(0) \sim \pm 0.3$ yr$^{-1}$, we again see stronger effects, with larger-amplitude librations. Thus, the character of the forced solution for a librating $\gamma$ is similar regardless of amplitude -- the effects are strongest for $\phi$ circulating at about twice the separatrix distance from exact resonance.
	
\begin{figure}
		\includegraphics[width=\columnwidth]{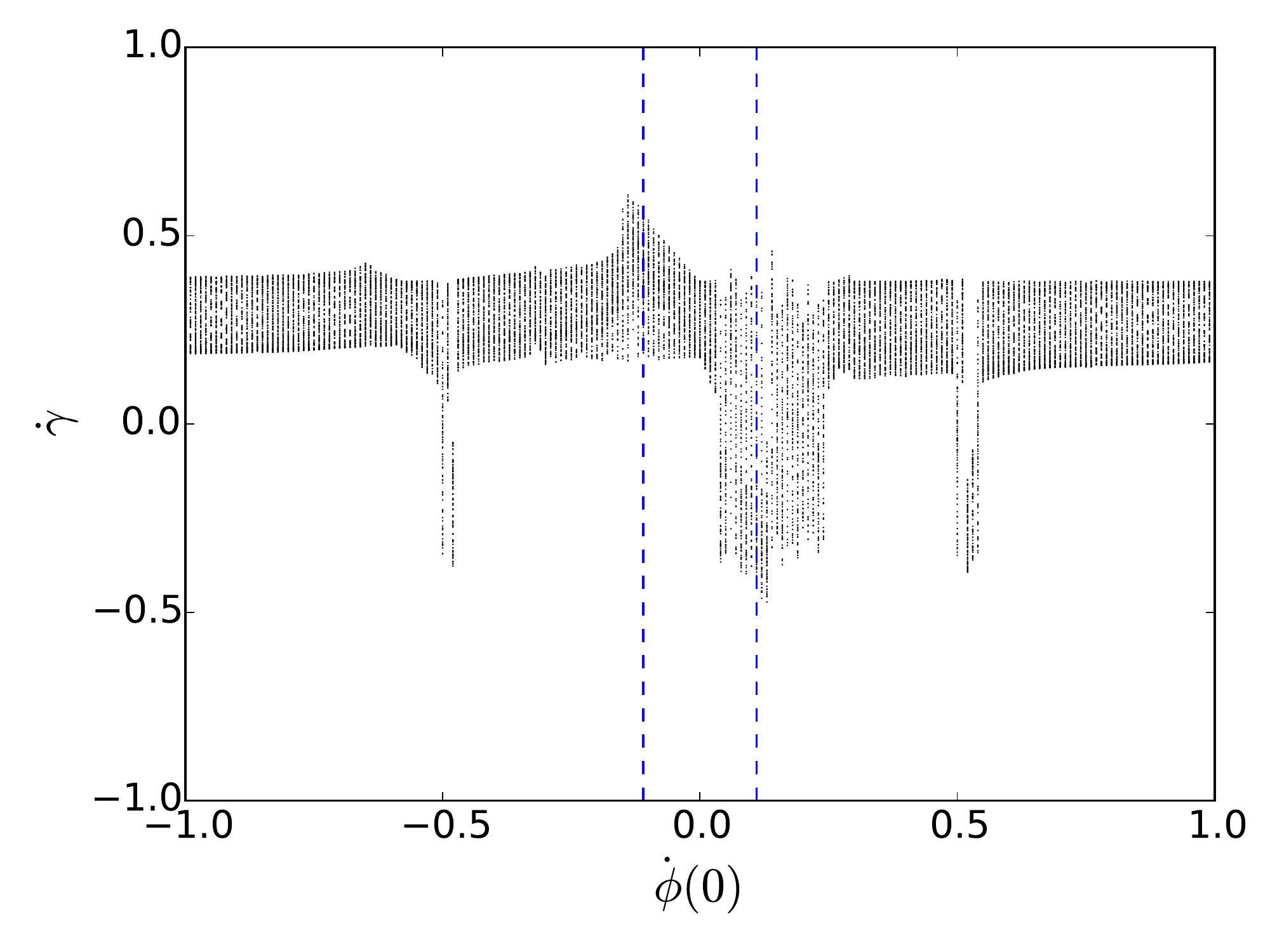}
		\caption{The points depict all $\dot{\gamma}$ (not just the surface of section) depending on the different initial forcing $\dot{\phi}$(0) among our simulations for our nominal K00255 values ($\beta = 0.11$) in the absence of damping $\epsilon = 0$ with $\dot{\gamma}(0) = 0.38$ yr$^{-1}$ (such that $\gamma$ is initially circulating). The vertical dashed lines indicate the seperatrix between libration and circulation of the external forcing resonant argument. Being in between these dashed lines is thus when the system is \textit{in} a mean-motion resonance, with the resonant angle $\phi$ librating instead of circulating. We note the asymmetry, finding more dramatic effects near the separatrix for positive $\dot{\phi}(0)$, where $\gamma$ evolution enters a wide chaotic regime.
			\label{gdvphid_k255_circ}} 
	\end{figure}

	The character of the behaviour for an initially circulating $\gamma$ is different. This is shown in Figure~\ref{gdvphid_k255_circ} by depicting values of $\dot{\gamma}$ against $\dot{\phi}(0)$, which presents a more useful picture than the Poincare projection in this specific case. Here, the effect of a librating $\phi$ is to enforce intermittent libration of $\gamma$ as well, but only if $\dot{\gamma}(0)$ and $\dot{\phi}(0)$ have the same sign. This is evident from the fact that Figure~\ref{gdvphid_k255_circ} shows far less symmetric behaviour about the origin than either of Figure~\ref{Amp} or Figure~\ref{Amp4}. In this case the influence of the perturber is stronger if the system is closer to resonance.
	
	\begin{figure}
		\includegraphics[width=\columnwidth]{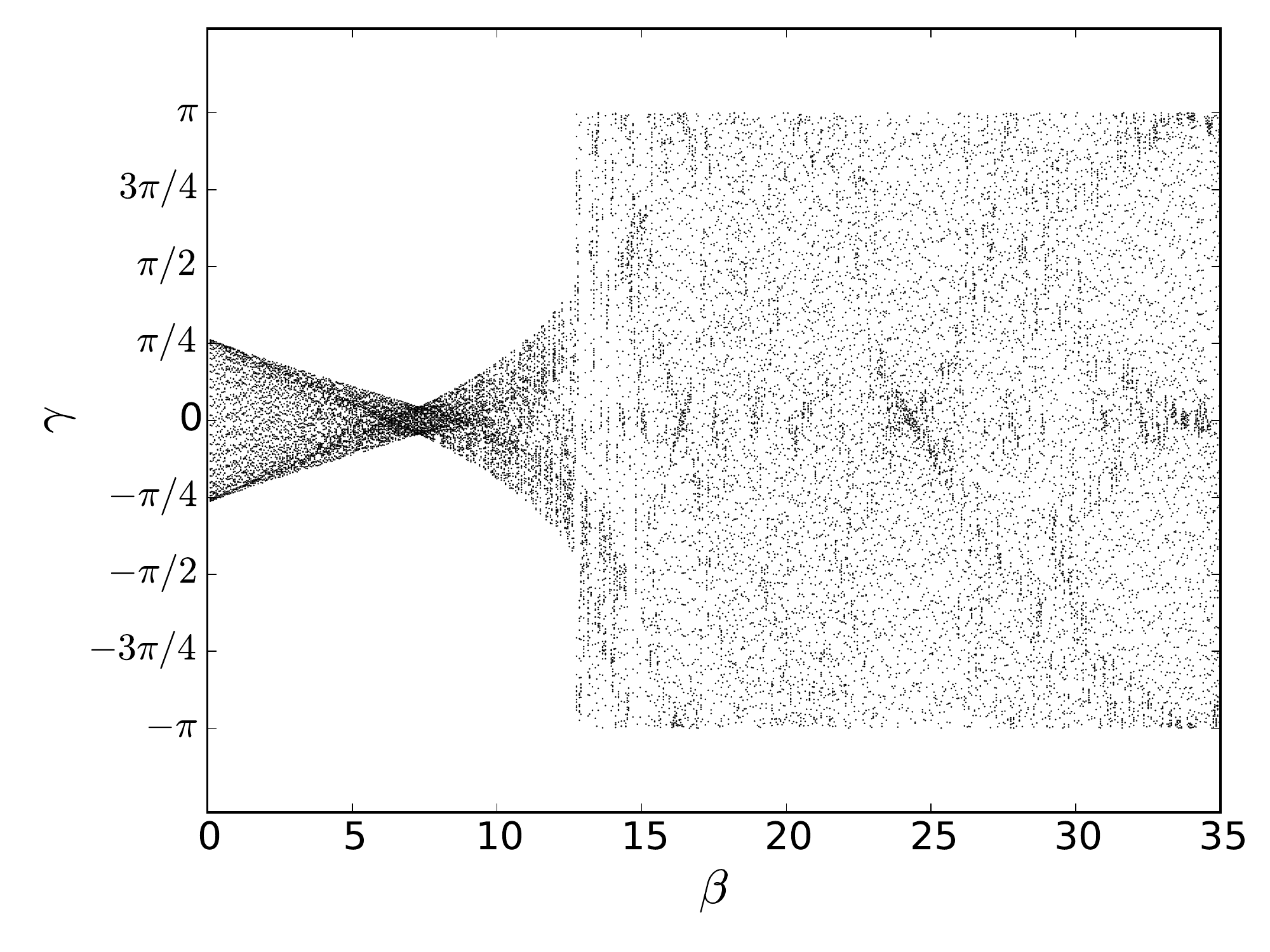}
		\caption[betavphid]{The points show the range of values for $\gamma$ in our simulations for each value of $\beta$, fixing $\dot{\phi}(0) = 2.0$ yr$^{-1}$ implied by observation of the K00255 system,  and with our nominal values for our K00255 system analogue. Here there is no damping term ($\epsilon = 0$), and $\dot{\gamma}(0) = 0.25$ yr$^{-1}$. We find that $\gamma$ does not begin to circulate until $\beta \gtrsim 12.5$, though we do find higher amplitude libration of $\gamma$ for lower values of $\beta$.
			\label{phidvbeta}}
	\end{figure}

	\begin{figure*}
		\includegraphics[width=\textwidth]{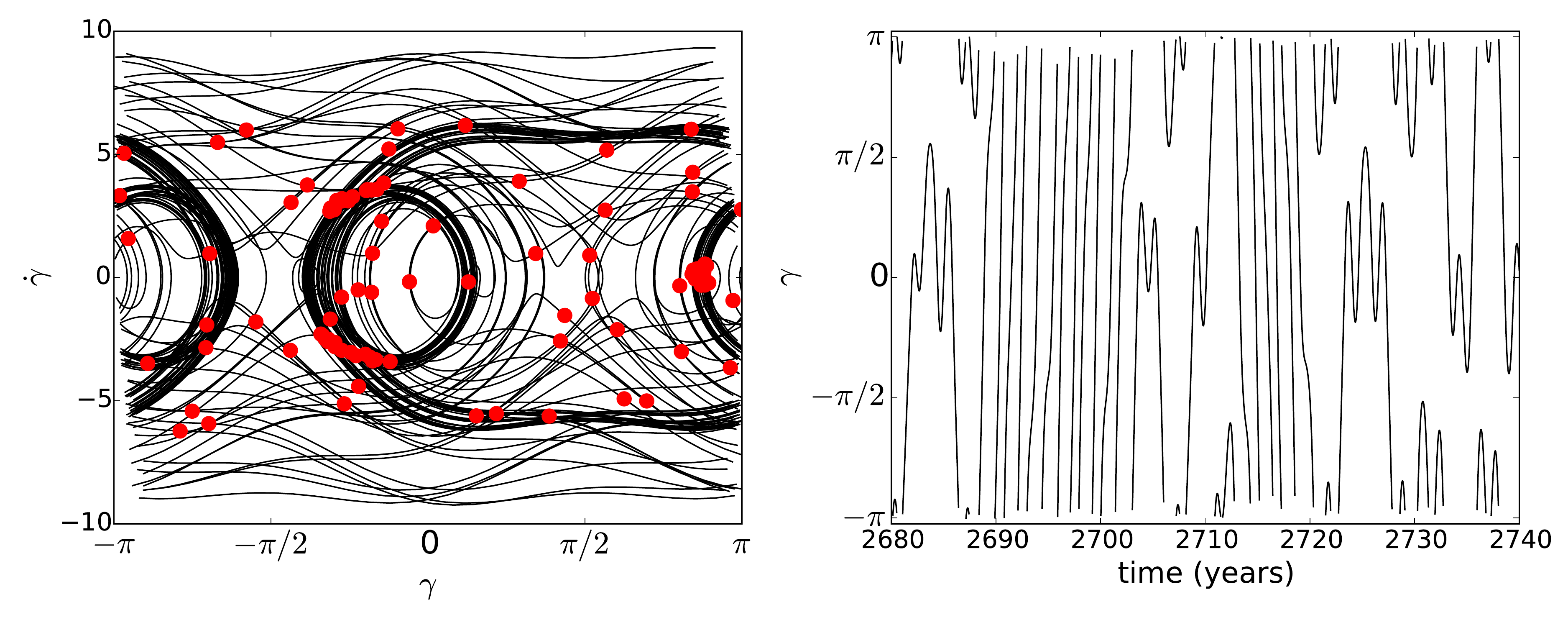}
		\caption[trappist_case_no_damp.pdf]{We show the evolution of $\gamma$ for nominal values of our TRAPPIST-1d system with $\beta = 1.0$ and no damping. We choose $\dot{\phi}(0) = 3.0$ yr$^{-1}$ based on the observed period ratio, and $\dot{\gamma}(0) = 0.3$ yr$^{-1}$. On the left we depict the evolution of $\gamma$ in $\dot{\gamma}$ -- $\gamma$ space, and the surface of section is depicted with red dots. On the right we show $\gamma$ as a function of time for a small snippet of time.
			\label{trappd}}
	\end{figure*}
	
	\begin{figure}
		\includegraphics[width=\columnwidth]{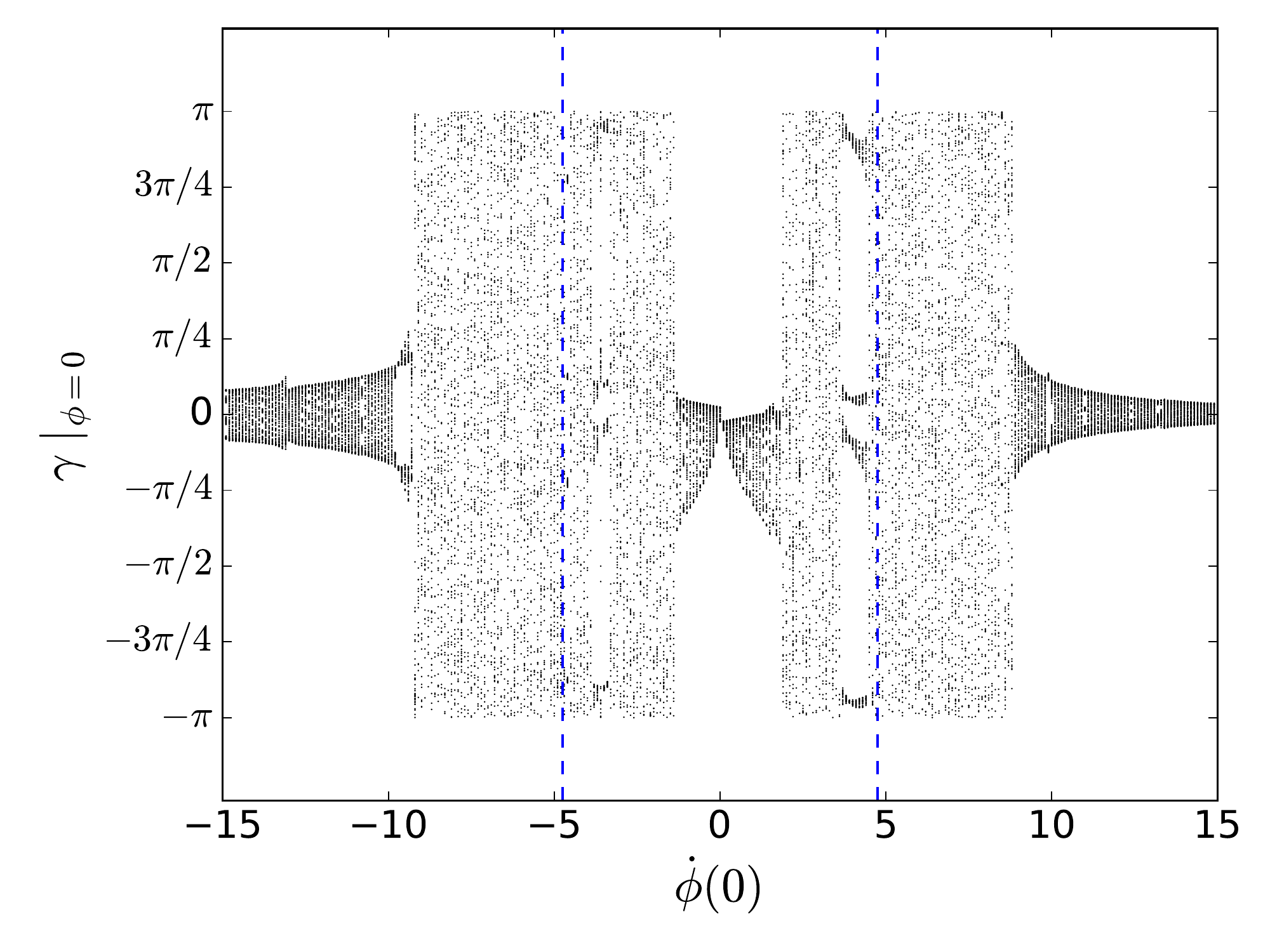}
		\caption{The points show the values of $\gamma$ in the surface of section defined by $\phi=0$, and $\dot{\phi}$ with the sign of its initial value, for each value of the initial forcing $\dot{\phi}(0)$ with our nominal values for the TRAPPIST-1d system with $\beta = 1.0$. Here there is no damping term, and $\dot{\gamma}(0) = 0.3$ yr$^{-1}$. The vertical dashed lines indicate the seperatrix between libration and circulation (to the far left and right) of the external forcing resonant argument. We can see that the behaviour is not symmetric about $\dot{\phi}(0)=0$, since the circulation of $\gamma $ imposes a preferred direction. We also observe a wide range of applicable $\dot{\phi}(0)$ as compared to the K00255 2:1 system.
			\label{Amp3}}
	\end{figure}
	
	\begin{figure}
		\includegraphics[width=\columnwidth]{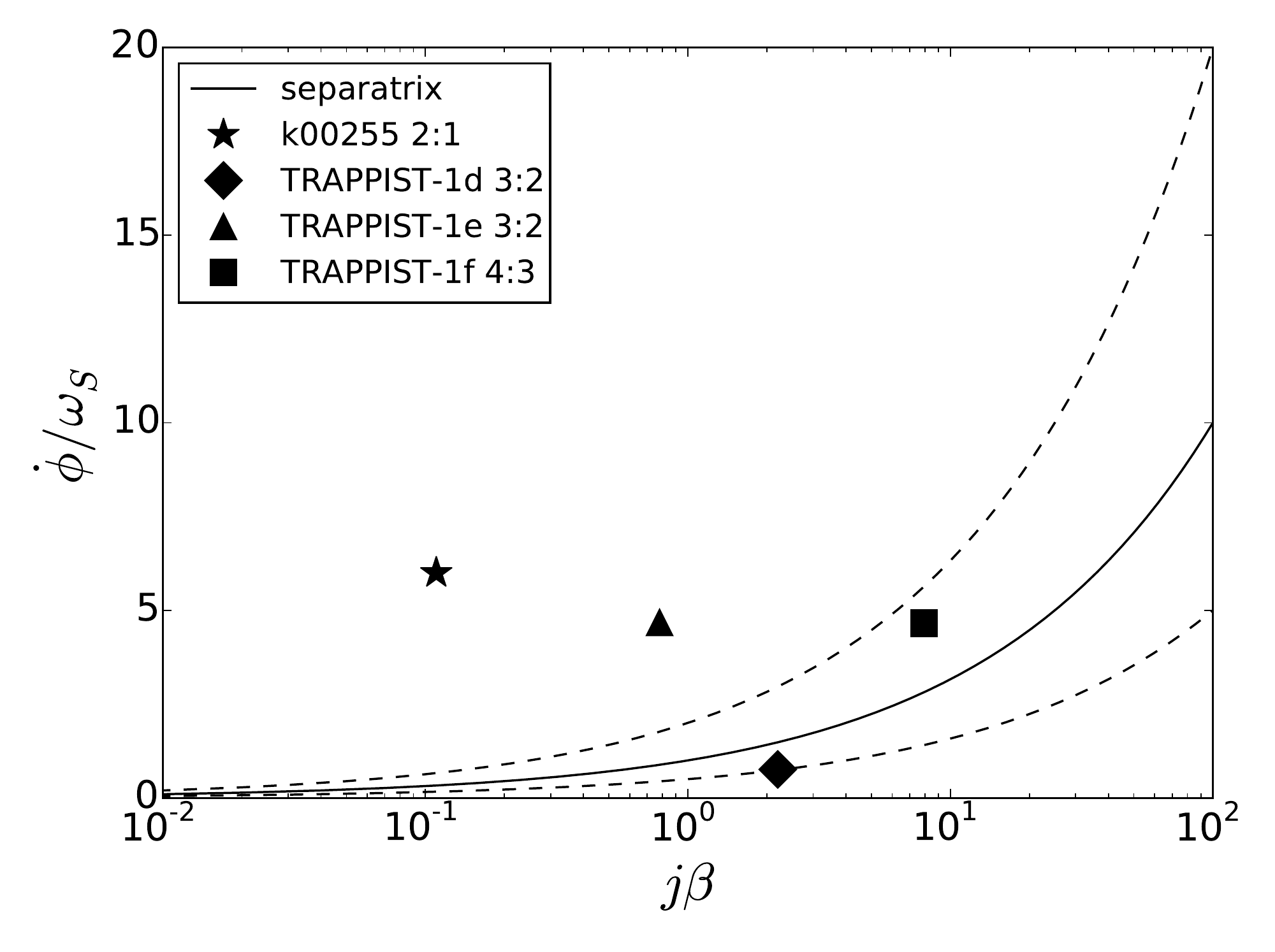}
		\caption[separatrix_v_beta.pdf]{We show how the location of the separatrix in the $\dot{\phi}$ -- $\phi$ system varies with $\beta$, depicting the value of $\dot{\phi}$ at the separatix when $\phi=0$. We also show the locations of the our K00255 and TRAPPIST-1d example systems, based on our normal parameters and implied $\dot{\phi}$ values from observation. The dashed lines depicted the location at twice and half the separatrix. Lying near or within the dashed lines imply higher potential for variance in $\gamma$. \label{phid_v_beta}}
	\end{figure}
	
	Overall, these estimates suggest that a system with the parameters of the K00255 system could experience episodes of chaotic spin evolution if close to a mean-motion resonance. We can infer what a likely $\dot{\phi}$ is from observation, with $\dot{\phi} = (j+1) n - j n'$ assuming $\bar{\omega} = 0$, and let $\dot{\phi}(0)$ equal this. For the observed K00255 period ratio of 2.024, we infer $\dot{\phi}(0)=2.0$ yr$^{-1}$, which is probably too distant from resonance to produce large effects. However, a less triaxial planet, with $(B-A)/C \sim 10^{-6}$, would lie within the chaotic zone at these periods.
	
	Given the uncertainty in the eccentricity, perturber mass, and especially $\left(B-A\right)/C$ in the K00255 system, it is also worth considering the sensitivity of the system to changes in $\beta$. If we fix $\dot{\phi}(0)=2.0$ yr$^{-1}$, consistent with the currently observed value based on period ratio, we can characterise the evolution of the spin for different values of $\beta$. For the case where $\dot{\gamma}(0)=0.25$ yr$^{-1}$, Figure~\ref{phidvbeta} shows how the spin-state varies with forcing amplitude $\beta$. We see that this system doesn't begin to undergo circulation, or chaotic motions, until $\beta > 12.5$. The character of the spin evolution is the same for larger values of $\beta$, except that the rotation gets progressively more rapid. From the threshold value of $\beta \sim 12.5$, we can estimate that a perturber mass of $\left (55 /e M_{\oplus} \right) |H(e)|$, with our fiducial $\left(B-A\right)/C = 2\times 10^{-5}$, is required for K00255.01 in order to get circulation. If we set a limit of $e<0.6$ to avoid orbit crossing, this gives us a strict lower limit of $17 M_{\oplus}$, although it is likely larger than this. For our nominal value of $e=0.05$, this would imply a larger mass of $555 M_{\oplus}$ to yield circulation in $\gamma$. These masses are much larger than the radius of the companion would suggest, and show that the K00255 system is likely too far from a mean-motion resonance for these effects to be very significant. If the planet is less triaxial, with $\left(B-A\right)/C = 2\times 10^{-6}$, these values change to a lower limit of $m' = 1.7 \ M_\oplus$, and  an implied mass of $m' = 56 \ M_\oplus$, still somewhat larger than the radius of the companion would imply. Nevertheless, for a similar system closer to a mean-motion resonance, circulation or larger amplitude oscillations in $\gamma$ is possible. We also note that the results also depend strongly on $\dot{\gamma}(0)$, which we will show in \S~ \ref{diss} could result in more dramatic effects, even when farther from resonance.
	
	Another potential application of this model is to the recently announced TRAPPIST-1 planetary system \citep{TRAP1}. In this case, there are several planets arranged in a chain, with several near commensurabilities. If we focus on the planet TRAPPIST-1d, which has the most Earth-like irradiation, and is found in a mean-motion resonance with planet TRAPPIST-1e, we find that $\beta \sim 1$ for the nominal choices of $e\sim 0.05$, $(B-A)/C \sim 2 \times 10^{-5}$ and the estimated mass for TRAPPIST-1e ($\sim 0.62 M_{\oplus}$). Other contributing factors to the larger $\beta$ are the fact that this is a 3:2 resonance, which increases $j$ and $C_r$, and that the host star is of very low mass ($\sim 0.08 M_{\odot}$).
	
	We infer $\dot{\phi}(0) = 3.0$ yr$^{-1}$ from the observed period ratio of the two planets if we assume $\dot{\phi}$ is currently at the maximum value in its oscillation. In Figure \ref{trappd}, we demonstrate our results for the evolution for $\gamma$ in this system if we choose $\dot{\gamma}(0) = 0.3$ yr$^{-1}$ (small-amplitude in natural libration). The red dots in the left-hand panel depict the Poincare points, recorded after each driving cycle, with their random placement illustrating the chaotic nature of that we happen to observe with these parameters. the right-hand panel shows a snapshot in a small time range, showing how $\gamma$ chaotically switches between libration about 0 or $\pi$ and circulation with either prograde or retrograde motion. For a broader picture, Figure~\ref{Amp3} shows the effect of the driving for different $\dot{\phi}(0)$ for this system, starting with $\dot{\gamma}(0)=0.3$ yr$^{-1}$. We see again a substantial effect around the separatrix ($\dot{\phi} (0) = 4.1$ yr$^{-1}$).
	
	It can be shown that the separatrix in $\phi$ occurs at $\dot{\phi} = \omega_S \sqrt{j \beta}$ and $\phi = 0$. With the knowledge that the largest effects occur near the separatrix of $\phi$, or around twice the separatrix in the 2:1 mean-motion resonance case, we can summarise the applicable range of parameters in Figure \ref{phid_v_beta}. Here we show how the separatrix of $\phi$ shifts as we vary $\beta$. For maximum effect, we would like to a find a system which lies roughly around the separatrix. We find that the K00255 system is likely too far from this range to yield interesting behavior, but the TRAPPIST-1d system lies squarely in this range, and indeed many of our simulations yield a circularised $\gamma$ near the implied $\phi$ separatrix based on observation. 
	
	The Planet TRAPPIST-1d is the one with the most earth-like insolation, but other planets in the system also reside close to mean motion resonances. Both TRAPPIST-1e and TRAPPIST-1f are also close to 3:2 and 4:3 commensurabilities respectively, and we estimate $\beta = 0.39$ and $\beta=2.65$ for these planets under the same assumptions as above. The large value of the latter is the result of both larger $j$ and $C_r$ and also a larger perturber mass. Our model can also potentially apply to TRAPPIST-1c, which lies near a second order (5:3) resonance. However, in this case $\omega_M^2$ (and thus $\beta$) acquires an extra factor of eccentricity, making the estimated value smaller ($\beta=0.04$). A further point to note for this system is that its extreme compactness also implies short tidal synchronisation {\em and} circularisation times. If we repeat the analysis above, we estimate synchronisation and circularisation times $\sim 3 \times 10^4$~years and $\sim 8 \times 10^6$~years, respectively, for TRAPPIST-1d. This would suggest that perhaps we should set eccentrity equal to zero for these systems. However, resonant chains with more than two planets can excite eccentricities through zeroth order resonances not available to the two planet case, as is known to occur in the Laplace resonance of the Galilean moons \citep{YP}. This combination can potentially lead to a non-negligible contribution to the planetary heat budget due to tidal dissipation \citep{BMG,VLBG,BSRL,DB}, which may shift the optimal conditions of habitability in this system from planet d to e or f. We review the influence of tidal dissipation from our model in appendix~\ref{Tides}. This system clearly warrants a more detailed investigation, but, for now, we simply assume that the more extended interactions generate eccentricity but that the spins can be described in the two planet formalism.
	
	\subsection{Limit Cycles including Dissipation \label{diss}}
	
	The preceding discussion focused on the interaction between the spin state and the external driving, in order to illustrate the kind of spin behaviours that are possible. In the true physical situation, we must also account for the damping effects of tides, which will drive the planet to synchronous rotation in the absence of an external forcing. When a forcing is present, we then expect the system to evolve to a finite amplitude limit cycle, or a stable circulation. In studies of harmonically driven damped pendula, the properties of this limit cycle can often be a sensitive function of the system parameters \citep{BB05}.
	
\begin{figure}
		\includegraphics[width=\columnwidth]{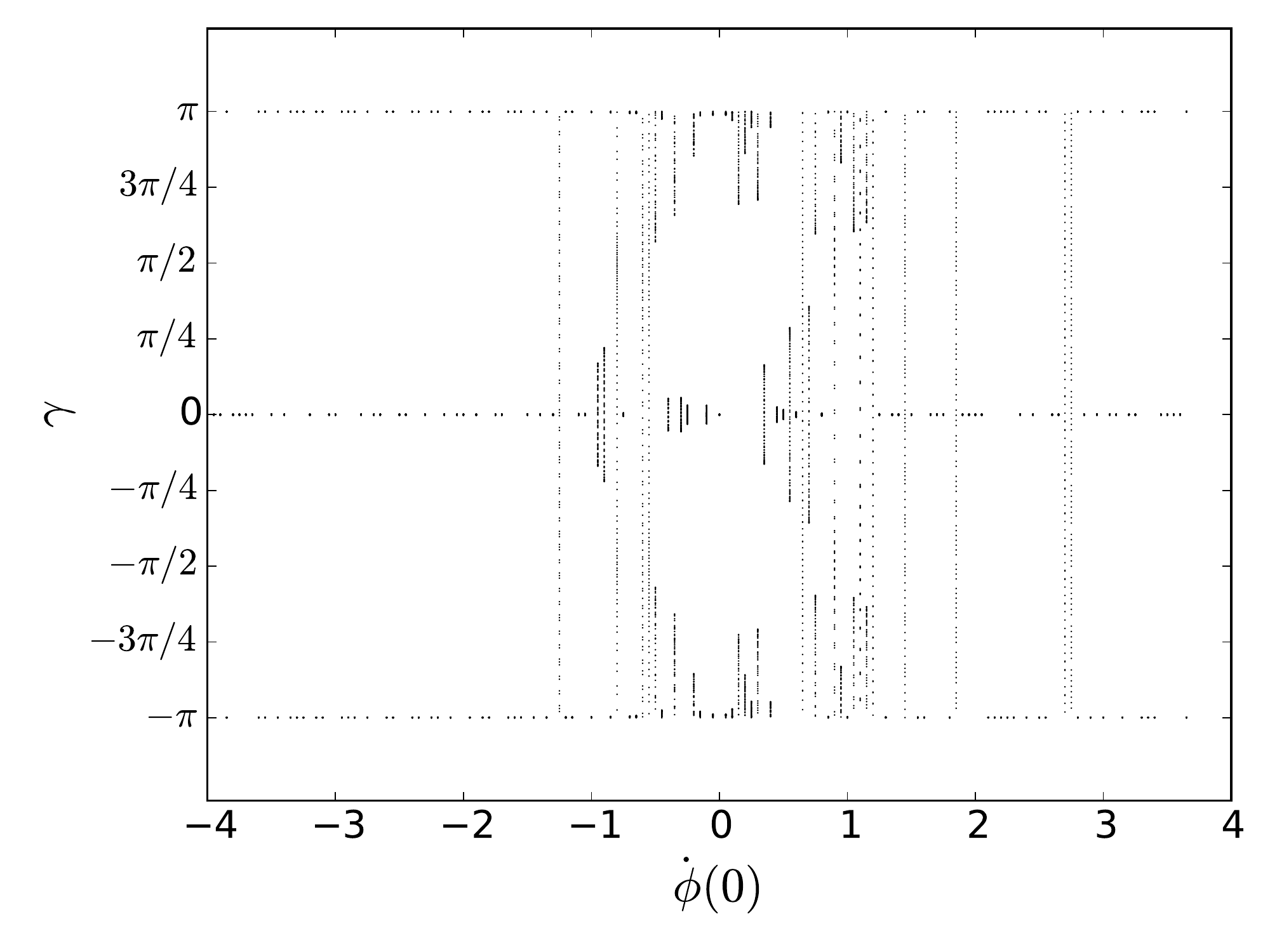}
		\caption{We show all values of $\gamma$ (not just the surface of section) executed by the limit cycle at late times as we vary initial conditions of $\dot{\phi}$ for nominal values of our K00255 system with $\beta = 0.11$ and dissipation factor $\epsilon = 6.0 \times 10^{-8}$.  We also assign a relatively high value of $\dot{\gamma}(0) = 1.5$ yr$^{-1}$. We find regions where $\gamma$ either circulates (where $\gamma$ spans the full range of $-\pi$ to $\pi$), or librates about $0$ or $\pi$, with final results being very sensitive to initial conditions.
			\label{gvphi_k_damp}}
	\end{figure}

	To examine the consequences of this, we now evolve equation~(\ref{eqn:full_motion}) for our nominal K00255 parameters (yielding $\beta=0.11$ and $\epsilon = 6 \times 10^{-8}$), assuming initially a large $\dot{\gamma}$, for a range of initial $\dot{\phi}(0)$, to examine the properties of the family of resulting limit cycles. Figure~\ref{gvphi_k_damp} shows our results with the points plotted depicting the full range (i.e. not just the Poincare section) of $\gamma$ probed by the solution at late times only ($\tau> 10^8$ yrs). Thus, these represent the final limit cycles and indeed show a great deal of sensitivity to the nature of the driving via $\dot{\phi}(0)$. The system can librate about either $\gamma=0$ or $\gamma=\pi$, and it can also circulate.  The final pattern of the limit cycle seems quite sensitive to the imposed conditions. As mentioned previously, the observed period ratio for the K00255 system suggests $\dot{\phi}(0)=2.0$ yr$^{-1}$. The solutions in this part of the diagram can exhibit limit cycle behaviour anywhere from complete circulation to low amplitude librations. We note that this is different than our solution for an initially librating $\gamma$, as shown in Figure \ref{Amp4}, without damping. This suggests that the initial conditions of the spin state, along with the strength of the forcing term, can dramatically alter the resulting limit cycle, even relatively far from the $\phi$ -- $\dot{\phi}$ separatrix and with subtle variations of these initial conditions, with higher $\dot{\gamma}(0)$ giving a larger range of interesting behaviors.
	
\begin{figure}
		\includegraphics[width=\columnwidth]{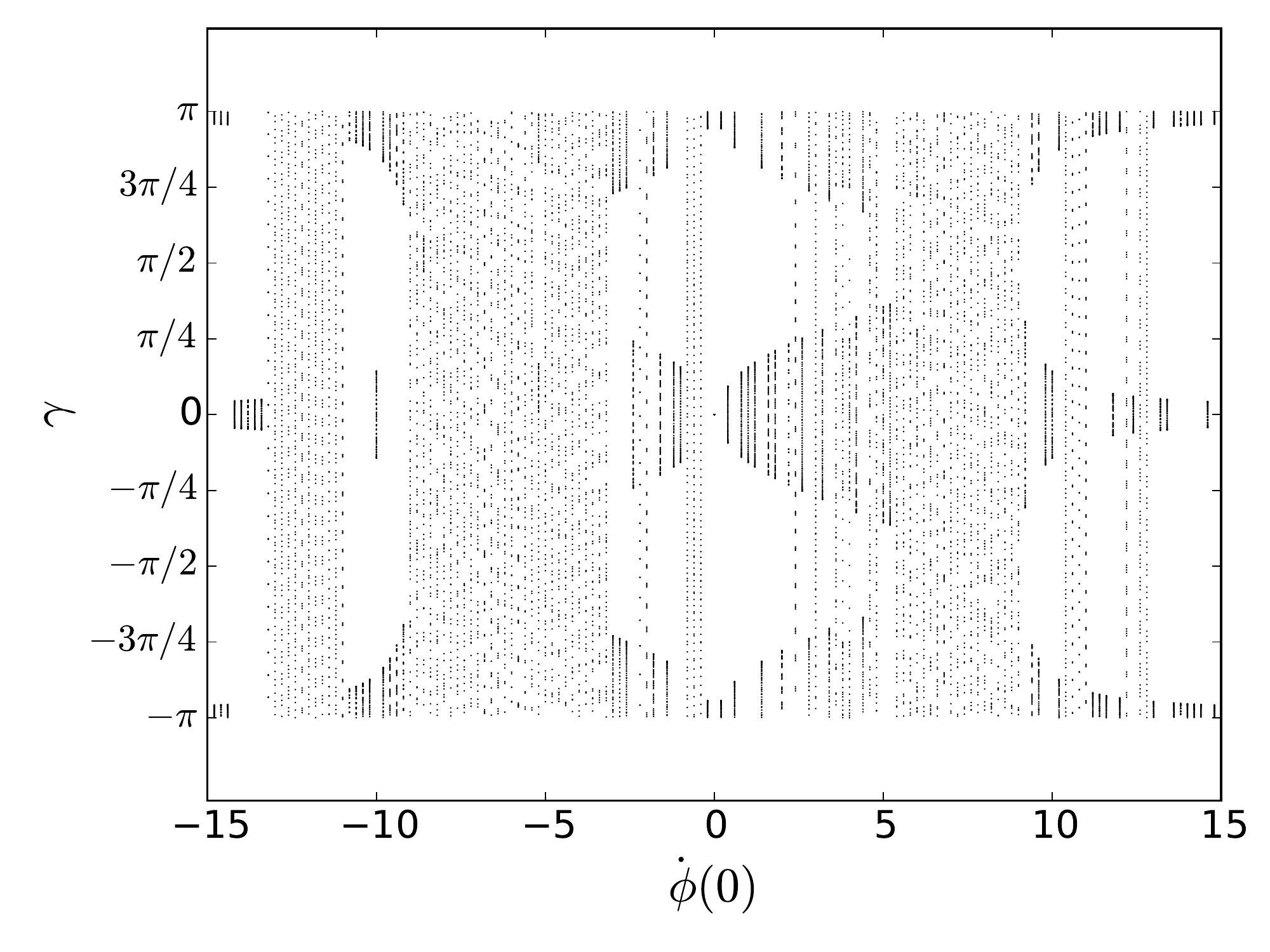}
		\caption[gvphid_trapp_g5.pdf]{The points show all values of $\gamma$ (not just the surface of section) executed by the limit cycle at late times under the influence of tidal damping, given the nominal values of the TRAPPIST-1d planetary system with $\beta = 1.0$ and $\epsilon = 8\times 10^{-3}$, and $\dot{\gamma}(0) = 5.0$ yr$^{-1}$. We see that, once again, there are large zones where the planet can undergo circulation in the limit cycle (where $\gamma$ spans the range from $-\pi$ to $\pi$), ensuring that the full planet surface receives some level of illumination.
			\label{Scan2}}
	\end{figure}

	Figure~\ref{Scan2} shows the same thing for the TRAPPIST-1d system. In this case, both the forcing and the damping are stronger, yielding $\beta=1.0$ and $\epsilon=8 \times 10^{-3}$ for our nominal system parameters, and we set $\dot{\gamma}(0) = 5.0$ yr$^{-1}$ (natural moderate-amplitude libration in $\gamma$). Once again we see similar behaviour, with more extensive regions where the limit cycle yields circulation. It is also interesting to note that $\dot{\phi}(0)=3.0$ yr$^{-1}$ is implied from observation, which yields a libration with semi-amplitude $\sim 42^{\circ}$, meaning that approximately $73\%$ of the planet gets at least some level of illumination. Furthermore, this is right on the edge of the regime where the limit cycle circulates. If $\dot{\phi}(0)$ were only slightly larger, or $(B-A)/C$ slightly smaller, then the entire planet could potentially also receive substantial illumination, and slight changes in parameters can yield wildly different results. Indeed, the evolution of $\phi$ is likely to be more complicated in this resonant chain system, and the planet may move between spin regimes. We intend to investigate this further in a forthcoming paper.
	
\begin{figure*}
		\includegraphics[width=\textwidth]{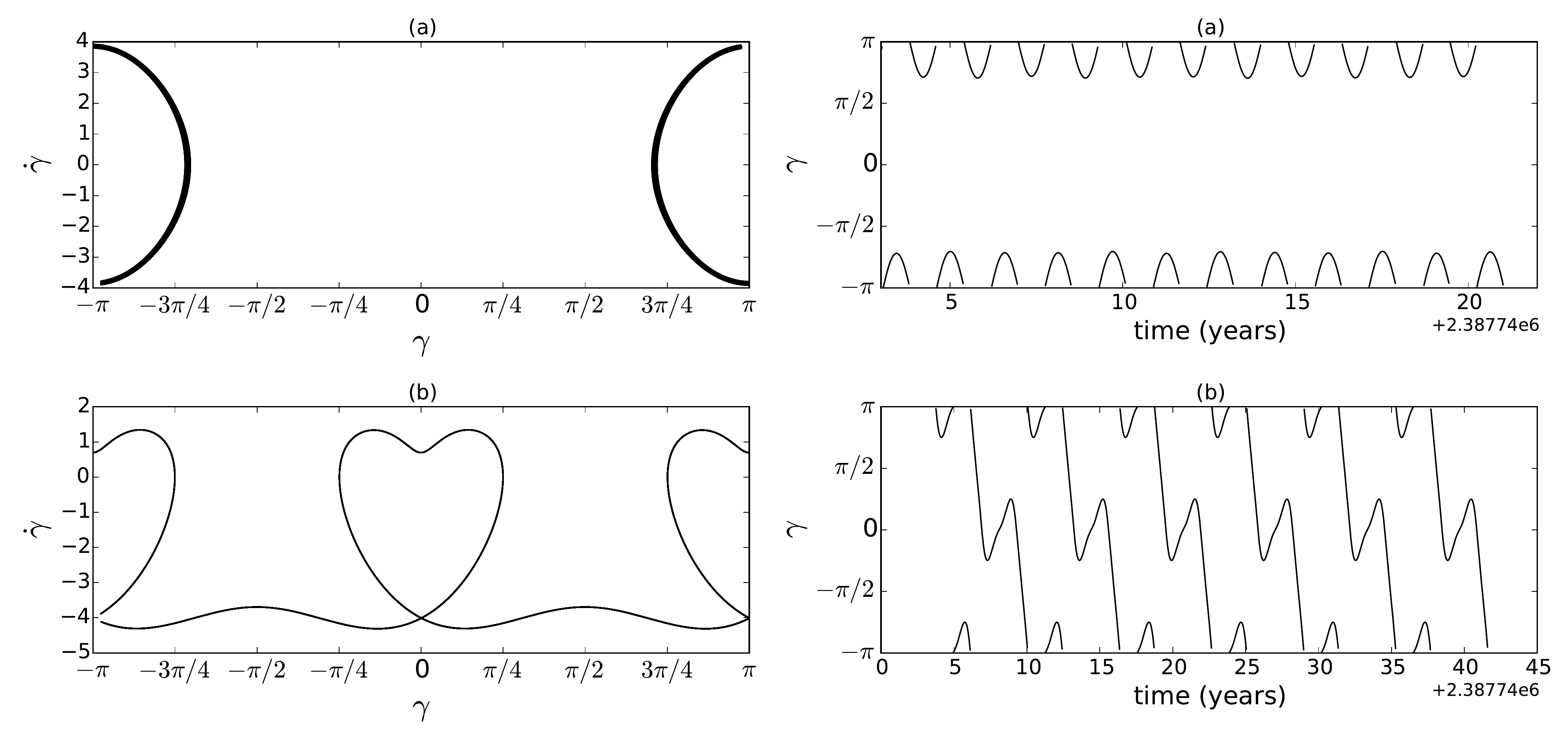}
		\caption[trapp_cases.pdf]{The curves show the evolution of $\gamma$ over time, presented in $\dot{\gamma}$ vs. $\gamma$ space in the left panels, and alternatively $\gamma$ vs. time on the right. Given our nominal values for the TRAPPIST-1d planetary system damping ($\epsilon = 8\times 10^{-3}$), $\beta = 1.0$, and $\dot{\gamma} = 5.0$ yr$^{-1}$, we compare two cases of slightly different $\dot{\phi}(0)$. In panels labeled (a), $\dot{\phi}(0) = 3.4$ yr$^{-1}$ and we observe libration about $\gamma = \pi$. In panels labeled (b), $\dot{\phi}(0) = 3.8$ yr$^{-1}$ and we observe elements of both libration and circulation. In both cases, we only see the behavior after the simulation has run for much longer than the dissipation timescale to allow the system to reach its limit cycle. The time after the start of the simulation is indicated in the bottom right of the time axes as $2.3877\times 10^6$ years.
			\label{trapp_cases}}
	\end{figure*}

	Figure~\ref{trapp_cases} shows two examples of the evolution to the limit cycle for the TRAPPIST-1d system near the separatrix of the $\phi$ -- $\dot{\phi}$ system. The top panels depict results with $\dot{\phi}(0)=3.4$ yr$^{-1}$ and $\dot{\gamma}(0)=5$ yr$^{-1}$, where we see that $\gamma$ damps down into a limit cycle librating about $\gamma=\pi$ with a semi-amplitude of $60^{\circ}$. The bottom panels of Figure~\ref{trapp_cases} shows the limit cycle with the same parameters except that $\dot{\phi}(0)=3.8$ yr$^{-1}$. In this case the limit cycle exhibits features of both circulation and libration -- in essence the circulation is interrupted by a two beat libration at both $\gamma = 0$ and $\gamma = \pi$, before resuming.
	
	\subsection{Timescales}
	
	In order to determine the effects on climate, we must also determine typical timescales on which the  spin states drift. The definition of $\dot{\gamma}$ implies the degree of mismatch between the spin of the planet and the orbital frequency, so the planet still rotates {\em almost} synchronously unless $\dot{\gamma}$ is quite large. However, a small finite mismatch means that the pattern of illumination will slowly drift over the surface of the planet, and the timescale over which it does this will determine the nature of the atmospheric response. If it occurs very slowly, the atmosphere should adjust to the change in illumination at each location, and should therefore present a similar atmospheric state to that of a synchronous planet, with the only likely variations due to whether the current dayside has more or less water present. On the other hand, if the drift is sufficiently rapid, the effect of the insolation is averaged over the surface of the planet. The most interesting cases are when the drift occurs on a timescale comparable to the atmospheric response time, as this can lead to nonlinear feedback.
	
	Specifically, we are interested in the time in which $\gamma$ spans the entire range within its limit cycle (i.e. the time for the substellar point to return to the same physical location on the planet and begin a new cycle). For circulating cases, this corresponds to the time in which $\gamma$ goes from $-\pi$ to $\pi$, after spanning all $\gamma$ in between. So we can say that this yields an effective length of a stellar day on the planet in these circulating cases. Meanwhile, for librating cases, it yields the time in which $\gamma$ makes a full oscillation with its maximum amplitude. We will define this as our relevant ``spin period.''
	
\begin{figure}
		\includegraphics[width=\columnwidth]{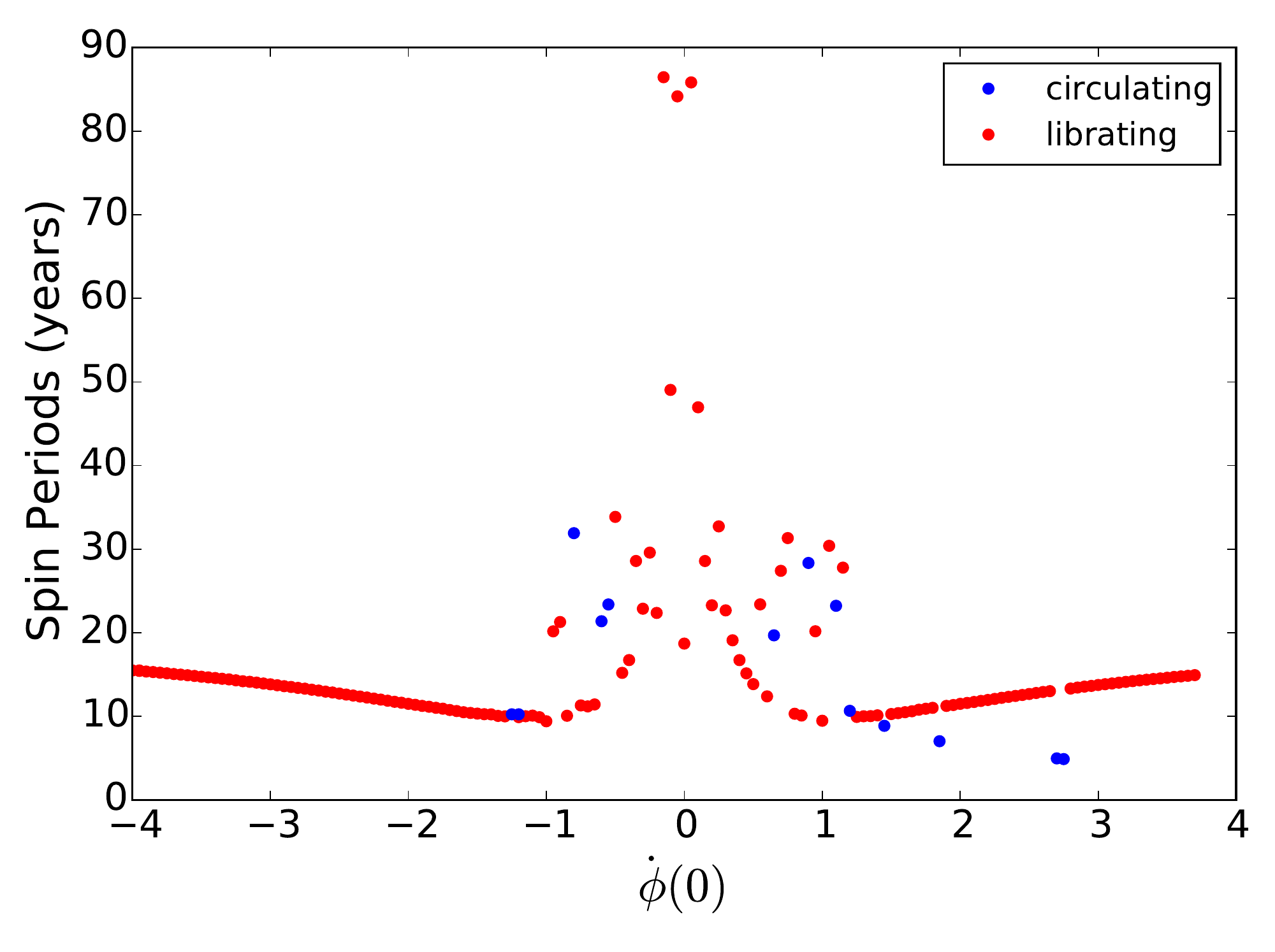}
		\caption[pvprox_k00255.pdf]{Here we show how the spin period varies as we change initial conditions in $\dot{\phi}(0)$ from the same simulations depicted in Figure \ref{gvphi_k_damp}. Each red dot depicts the values for simulations which yield librating spin states for $\gamma$, while blue points depict simulations which yield circulating $\gamma$ states. We observe that this system yields mostly librating states, though a few circulating cases exist, notably around the implied $\dot{\phi}(0) = 2.0$ yr$^{-1}$ from observation. These spin period timescales, however, are on the order of decades. \label{pvprox_k00255}}
	\end{figure}

	In the case of our K00255 analogue, the large scale librations and circulations shown in Figure~\ref{gvphi_k_damp} have characteristic timescales $\sim $10--25~years. We summarise this in Figure \ref{pvprox_k00255}, where we show what our relevant spin period is for each simulation, where red dots depict drift periods for librating cases, and blue depicts circulating cases. 
	
\begin{figure}
		\includegraphics[width=\columnwidth]{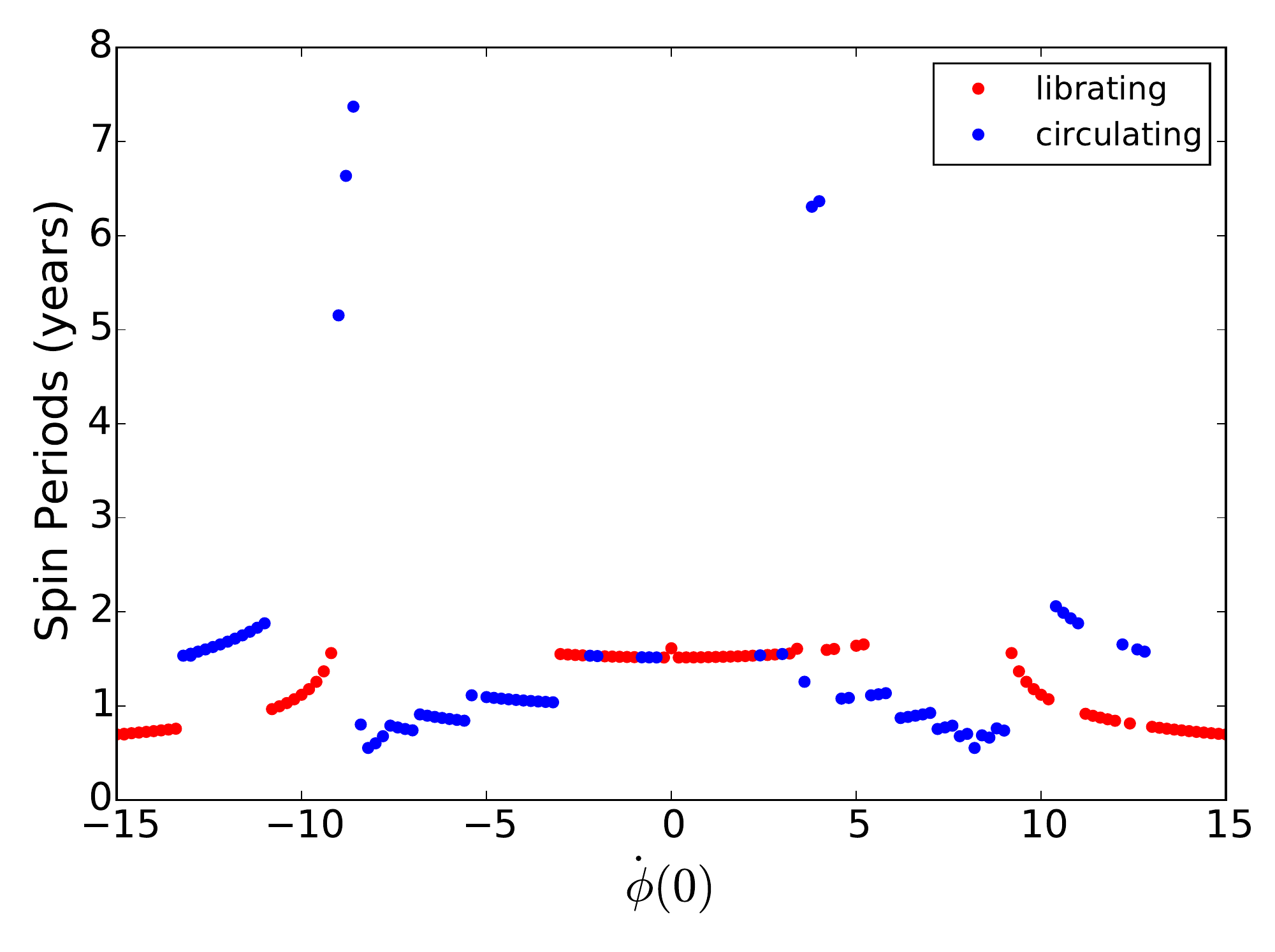}
		\caption[pvprox_trappist.pdf]{Here we show how the spin period varies as we change initial conditions in $\dot{\phi}(0)$ from the same simulations depicted in Figure \ref{Scan2}. Each red dot depicts the values for simulations which yield librating spin states for $\gamma$, while blue points depict simulations which yield circulating $\gamma$ states. We note that abundance of simulations which yielded circulating spin states, with typical timescales on the order of a year. \label{pvprox_trappist}}
	\end{figure}

	If we consider the two limit cycles for TRAPPIST-1d shown in Figure~\ref{trapp_cases}, the period for the libration for the case in the top panel is 1.5~years. This suggests that portions of a planet in such a configuration could experience variations in levels of illumination not unlike the seasonal variations experienced on Earth because of the obliquity of Earth's spin. In the case of the hybrid libration-circulation limit cycle shown on the bottom panels of Figure \ref{trapp_cases}, the period of the libration episodes is similar, and the period of a full cycle, including the circulation, is 6~years. We can again summarise this in Figure \ref{pvprox_trappist}, with red dots depicting librating cases, and blue dots circulating cases. This shows that the TRAPPIST1-d system has much shorter spin periods than our K00255 system analogue, with many periods on the order of just a year, as well as having far more circulating cases.
	
	\begin{figure}
		\includegraphics[width=\columnwidth]{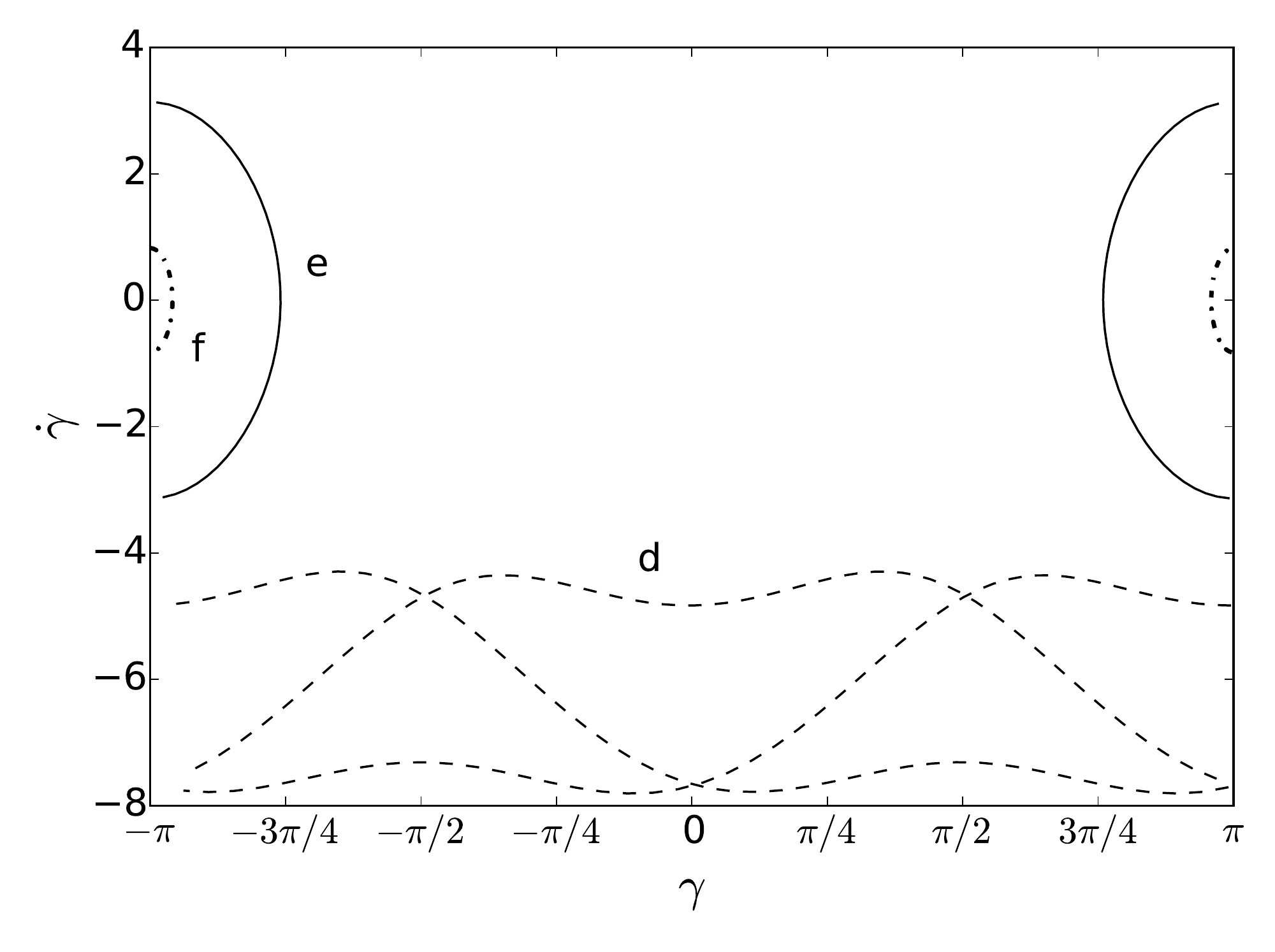}
		\caption[3cases.pdf]{We show the limit-cycle solution for 3 different planets inspired by the TRAPPIST-1 system (TRAPPIST-1 d, e, and f), considering only the effects of the immediate outer companion for each and using initial conditions in $\dot{\phi}(0)$ that would be suggested from current observation of the respective period-ratios. The dashed line depicts the solution for TRAPPIST-1d, which shows a circulation of the spin state. Meanwhile, TRAPPIST-1e, depicted by the solid line, has reached a moderate-amplitude librating state about $\pi$. TRAPPIST-1f, depicted by the dashed-dotted line, librates about a much $\pi$ at a much smaller amplitude, and thus represents a nearly perfect synchronous spin state. \label{3cases}}
	\end{figure}
	
	\begin{figure}
		\includegraphics[width=\columnwidth]{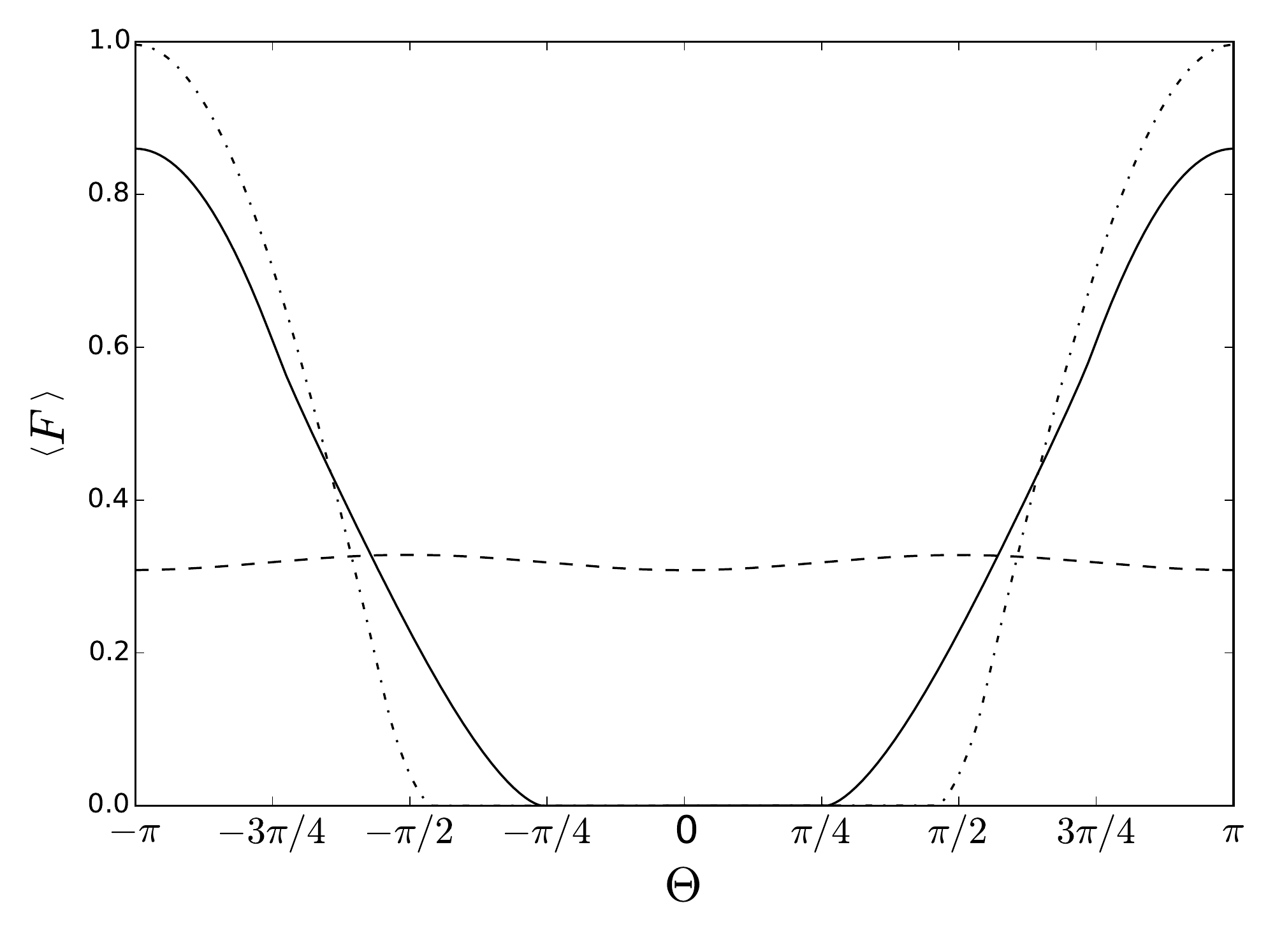}
		\caption[3cases.pdf]{Here we show the average flux received, normalized to the flux at the substellar point, at each longitude $\Theta$ of the planets depicted in Figure\ref{3cases} over time (with corresponding line styles). We see that, in the circulating case (planet d), all longitudes receive approximately the same amount of stellar radiation over time, with some slight variations. The other two are both librating cases, with planet f representing a nearly perfect synchronous spin state about $\pi$.  We observe a widening for the higher-amplitude case (planet e) as compared to the more perfectly synchronous planet f, with a higher portion of the planet receiving at least some stellar coverage. \label{instell}}
	\end{figure}
	
	Overall, the large scale librations and circulations in the limit cycles of equation~(\ref{eqn:full_motion}) have characteristic timescales in the range $\Delta \tau \sim 1$--25 years, which can imply different real-world timescales depending on the triaxiality and orbital period, but which are approximately decades for Earth-like planets in the habitable zones of M0 stars, and of order years for those in the habitable zones of very low mass stars, like TRAPPIST-1.
	
	\section{Discussion}
	\label{Discuss}
	
	Our model demonstrates that the presence of a companion in, or near, a mean motion resonance can perturb a planets spin sufficiently, even in the presence of tidal dissipation, to place it in a non-synchronous limit cycle. This can result in substantial drift, and even circulation, of the planetary spin. A consequence of this is that  a greater fraction of the planetary surface can receive some level of illumination, relative to that expected from a planet tidally locked to a synchronous spin. Furthermore, the timescales for this drift are approximately years to decades, depending on the particular configuration. This could have profound implications for the climate, and thus the habitability, of such a planet. The response of a planetary atmosphere and surface conditions to variations in level of illumination is very complex and an entire subject of study in itself. Nevertheless, we can anticipate the broadest atmospheric consequences of our new spin dynamics by studying the one-dimensional energy balance models used to estimate the breadth of stellar habitable zones (e.g. \cite{KWR, K13}). In particular, if we consider the models of \cite{SMS}, the atmospheric thermal response times for Earth-like planets range from years to decades, depending on the level of surface water (more water implies more thermal inertia and longer response times). This is comparable to the characteristic timescales estimated above for variations in the level of illumination, which implies that the atmospheric dynamics could be quite complex, as the conditions change on similar timescales to the response.	In the context of exoplanet habitability then, our results suggest one more avenue that may allow a planet to avoid the consequences of tidal synchronism close to low mass stars \citep{KWR}. Indeed, our model need not be exclusive of other proposals to avoid this fate. The formation of thick clouds near the substellar point \citep{YBFA} will presumably still be a part of the evolving climate dynamics, and the operation of strong atmospheric tides \citep{LWMM} may also operate, introducing an additional torque component into the spin dynamics that might also be worthy of consideration in the future. We began our consideration of this problem motivated by the anticipation of future discoveries of planetary systems around M dwarfs, and use the K00255 system as a prototype for the kinds of host stars likely to dominate magnitude limited samples of this type. However, the discovery of the planetary system around the very low mass ($\sim 0.08 M_{\odot}$) star TRAPPIST-1 illustrates that such models have a potentially very broad scope, and the presence of several near-commensurabilities in that system suggest that the configurations considered here are not  rare. The fact that the drift timescales discussed above scale relative to the orbital period also suggests that the particular atmospheric responses may vary with stellar host mass too, with the habitable planets around the lowest mass stars like TRAPPIST-1 experiencing variations quite similar to the seasonal variations on Earth. In near resonant chains like the TRAPPIST-1 system, the variation in the spin dynamics may have an influence on the relative habitability of each of the systems as well. Figure~\ref{3cases} shows the limit cycle for the three planets TRAPPIST-1d, TRAPPIST-1e and TRAPPIST-1f, that lie near first order resonances. In each case we have considered only the effects from the external neighbour, even though the inner companion may also provide non-negligible perturbations. In each case, there is a reason to suggest that the external perturber is more significant (the coupling between c-d is second order, and the external perturber is more massive in the case of planets e and f). We have performed this calculation assuming the nominal triaxiality and $Q_{\oplus}$ for each planet. We see that TRAPPIST-1e executes moderate libration under these conditions, and TRAPPIST-1f a much smaller libration, both about $\gamma = \pi$. On the other hand, TRAPPIST-1d has a limit cycle that circulates and features an intermittent, wide-angle libration as well. The characteristic timescale for the full cycle is $\sim$ 1.1~years in this case. This demonstrates that planets in the same system can occupy different spin states, with some being essentially synchronised, while others experience a more distributed irradiation. Although the arrangement in Figure~\ref{3cases} counter-intuitively suggests increased synchronism away from the star, this more has to do with our exact choices in initial conditions, with strong sensitivity to $\dot{\gamma}(0)$ and especially $\dot{\phi}(0)$ within these regimes that could result in significantly different limit cycles. The general effect is typically regulated by the value of $\beta$, but the solutions could still change dramatically within certain regimes if we adjust $\dot{\phi}(0)$ or $\dot{\gamma}(0)$, as can be seen in Figures \ref{gvphi_k_damp} and \ref{Scan2}. 
	
	Such spin states have potentially both direct and indirect observational consequences. The most obvious indirect consequence is simply the fact that spreading the stellar illumination over more of the surface area will help to make a planet more temperate and increase the fraction of the planet that is habitable. Figure~\ref{instell} demonstrates this by comparing the average flux received at each point on a planet for the limit cycles shown in Figure~\ref{3cases}. In the case of libration for TRAPPIST-1d, we see a moderate increase in the average value of the flux and in the surface area irradiated. For the circulating case, the differences are much more dramatic, with a substantial averaging observed. More direct observational consequences may emerge if the observing technology reaches the point of identifying signals from surface features (e.g. \cite{FST, KF11, CVA, C13}), because then one could potentially identify the years-to-decades drift on surface features over the course of many orbits.
	
	\section{Concluding Remarks}
	
	As observations uncover more planets around other stars, it is becoming clear that many such systems contain multiple planets, orbiting in compact configurations with substantial mutual gravitational interaction. Such interactions are particularly strong when the mean motions of neighbouring planets are approximately commensurate with low order rational ratios. In that event, the induced periodic variation in mean motion is often enough to prevent the planetary spin from achieving a stable equilibrium, even under the influence of strong tidal damping. Contemporaneously with our work, \cite{Delisle} have presented an analysis of a similar problem, focussed on near-resonant systems discovered by Kepler, for which significant transit timing variations are observed. They identify the same qualitative behaviour as we do, namely the existence of asynchronous and possibly chaotic spin equilibria. As shown by our work above, the application of such considerations to planets in the habitable zones of low mass stars offers a mechanism to alleviate the long-standing problem of asymmetric illumination of such planets.
	
	We have examined the dynamics of the planetary spin and identified a variety of limit cycle behaviour, including libration, circulation and combinations of both. Furthermore, the characteristic timescales for the stellar illumination to drift over the planet appear to be from years to decades, which is comparable to the thermal inertia of terrestrial planet atmospheres. This promises a variety of interesting feedback behaviour and suggests that the climates of habitable planets around M dwarfs may be far more dynamically rich than the traditional synchronous rotation often assumed. Unfortunately the nature of the limit cycle solutions are quite sensitive to the parameters of the problem, some of which are difficult to constrain directly. Nevertheless, our results suggest that such systems may reward  efforts to characterise their atmospheric properties directly.
	
	Such considerations are likely to become ever more important in the forthcoming years, as the observational sample continues to grow through both ground-based and space based efforts such as TRAPPIST and TESS. The recent announcement of the discovery of the TRAPPIST-1 planetary system is particularly encouraging in this regard, since it demonstrates the reality of the mean motion resonant configurations discussed in this paper.
	
	\bibliographystyle{mnras}
	\bibliography{refs}
	
	\appendix
	
	\section{The Shift of the Habitable Zone due to Tidal Heating}
	\label{Tides}
	The traditional definition of a habitable planet is one that can maintain a surface temperature sufficient to allow for the long-term existence of liquid water at the surface. The particular conditions that allow this to occur are the subject of detailed study (see \cite{KWR} for a discussion of the standard picture), but a simple estimate is often made by scaling relative to the amount of insolation that the Earth receives from the Sun. We shall adopt heating rates with a factor of two of Terrestrial Insolation as our simple definition of Habitability.
	
	For most stars, heating is equivalent to instellation -- the amount of irradiation received from the central star. However, for very low mass stars, habitable planets may be quite close to the star, and tidal forces may generate significant heat if there is a non-zero eccentricity \citep{BMG,VLBG,BSRL,DB}. Since our model invokes a non-zero eccentricity to provide a resonant perturbation, we estimate here the contribution to total heating that arises from out tidal model.
	
	If we calculate tidal heating due to circularisation from equation~(\ref{tcirc}), within the formalism of \citep{H10}, and add this to the heating received from instellation (assuming the energy is redistributed evenly across the surface of the planet), we obtain an expression for the surface temperature of a planet orbiting a $0.08 M_{\odot}$ star
	
	\begin{equation}
	T^4 = 245^4 \left( \frac{a}{0.03 {\rm AU}} \right)^{-2} + 223^4 \left( \frac{e}{0.05} \right)^2
	\left( \frac{R_p}{R_{\oplus}} \right)^8 \left( \frac{a}{0.03 {\rm AU}} \right)^{-9}. \label{Tbalance}
	\end{equation}
	
\begin{figure}
		\includegraphics[width=\columnwidth]{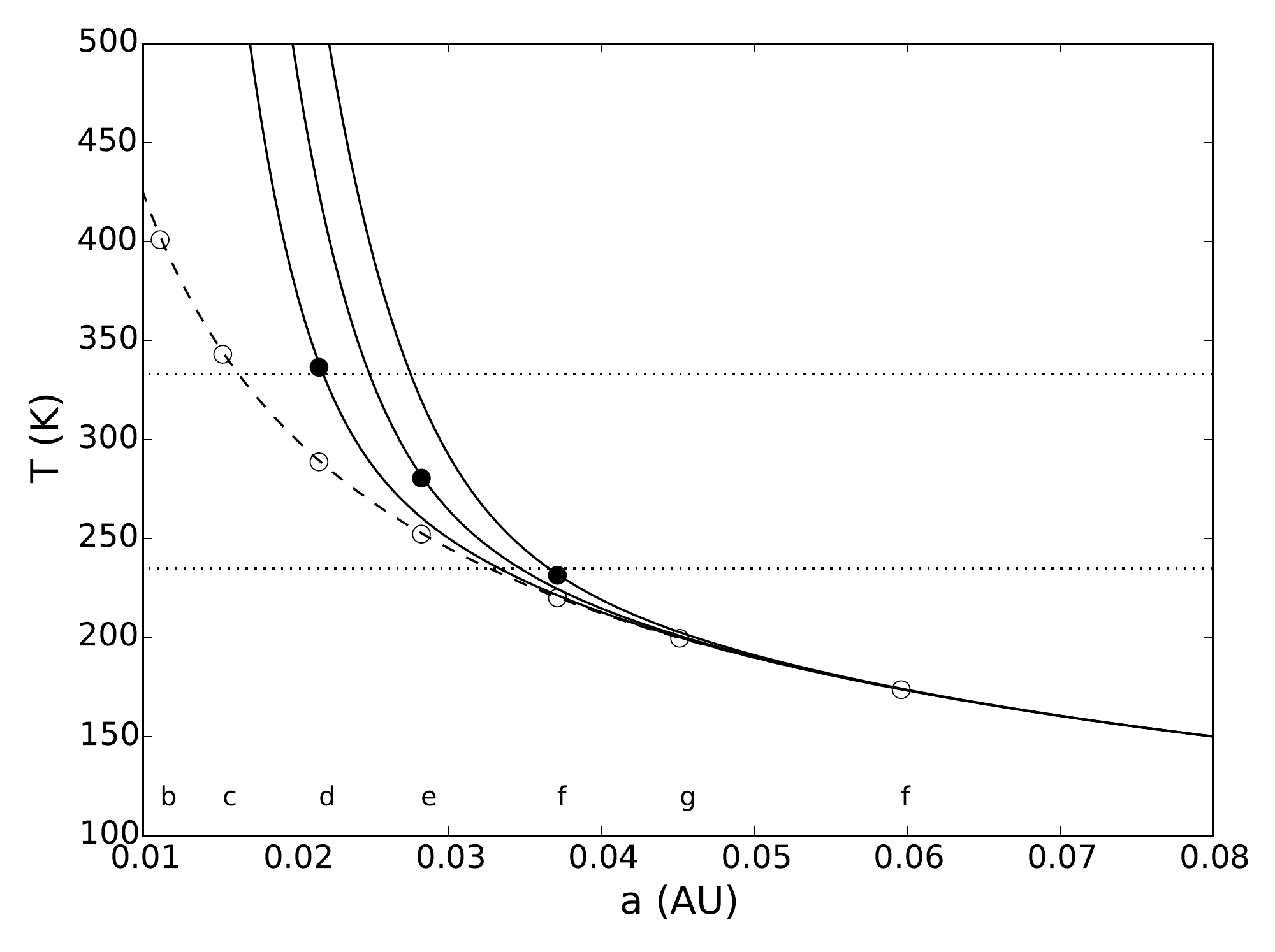}
		\caption[temp.pdf]{The dashed curve shows the estimated surface temperature maintained by stellar irradiation alone, assuming a central host star of mass $0.08 M_{\odot}$. The open circles therefore show the estimated surface temperatures for the TRAPPIST-1 planets, as labelled at the bottom. The solid curves show the revised surface temperatures including the internal contribution from tidal dissipation using our model. The three curves are calculated assuming planetary radii $0.77 R_{\oplus}$, $0.92 R_{\oplus}$ and $1.05 R_{\oplus}$ respectively. The solid circles show the revised estimates for the temperatures of each TRAPPIST-1 planet, and the horizontal dotted lines bracket a total heating rate within a factor of 2 of the irradiation received by the Earth from the Sun. We infer that TRAPPIST-1d and TRAPPIST-1e are still habitable, and that TRAPPIST-1f may be more likely to be habitable with the tidal heating contribution.
			\label{TrapHab}}
	\end{figure}

	The second term represents the tidal heating, and depends strongly on the planetary radius andalso on the eccentricity. The consequences for planetary temperature are shown graphically in Figure~\ref{TrapHab}, which shows curves for the radii of each of the three planets TRAPPIST-1d, TRAPPIST-1e and TRAPPIST-1f. These are the most likely to be habitable, and the tidal heating doesn't change this conclusion significantly, and actually improves the case for TRAPPIST-1f. As these are the three planets most likely to exhibit interesting spin dynamics, the contribution of tidal heating will add another dimension to the atmospheric dynamics.		
	
	Equation~(\ref{Tbalance}) and Figure~\ref{TrapHab} are based on the orbital energy dissipated due to tidal circularisation. However, in this paper, we have focussed more on the dissipation of spin. Is dissipation of spin energy not of interest? In the same formalism, we can calculate the energy release due to spin dissipation relative to orbital circularisation, as
	
	\begin{equation}
	\frac{\dot{E}_{spin}}{\dot{E}_{circ}} \sim \frac{2}{19} \frac{\dot{\theta}}{n} 
	\left( 1 - \frac{\dot{\theta}}{n} \right) \sim 0.23 
	\left( \frac{e}{0.05} \right)^{-2} \dot{\gamma}
	\end{equation}
	
	where we have normalised $\dot{\gamma}$ by $\omega_s/\sqrt{2}$ as in the rest of the paper, assuming $(B-A)/C = 2 \times 10^{-5}$. Therefore, we conclude spin energy is a minority contributor when considering the slow drifts relative to synchronism of interest here.
	
\end{document}